# On the Non-Markovian Navier-Stokes Framework for Turbulence Modeling- A Preliminary Analysis

Siamak Kazemzadeh Hannani and Ehsan Ghaderi

Department of Mechanical Engineering, Sharif University of Technology

Tehran, Iran

## Abstract

This study explores a formulation of the Navier-Stokes equations (NSE) using fractional calculus in modeling turbulence. By generalizing the stress-strain constitutive relation to incorporate nonlocal spatial interactions ($(-\Delta)^{1/3}$) and memory effects ($\partial_t^{1/2}$), we redefine a fractional Navier-Stokes equation (fNSE). Regarding the inertial-range scaling, the fractional Laplacian of order $\alpha=1/3$ and time-fractional derivative ($\beta=1/2$) capture non-Markovian energy transfer. The one-dimensional advection-diffusion equation, for the purpose of initial validation and Burgers' non-linear equation for the energy spectrum behavior are employed to investigate numerically the fNSE formulation. Depending on the strength of the fractional term we showed that, for the Burgers' equation, the numerical solution exhibits three distinct regimes: (1) early-time spectral broadening as nonlinearity transfers energy to smaller scales, (2) intermediate shock formation with characteristic $k^{-2}$ scaling, and (3) late-time fractional equilibrium where the anti-diffusive term sustains a shallower $k^{-4/3}$ inertial range - a symbol of fractional turbulence. Moreover, the transient one-dimensional heat equation and the Caputo derivative embedded Burgers' equations are solved, demonstrating the solution behavior regarding temporal memory effects. To simulate turbulent kinetic energy decay, we numerically solve the incompressible NSE using a pseudo-spectral method in a 3D periodic domain, demonstrating the fNSE's solution behavior. Key unresolved challenges include: Enforcing boundary conditions in fractional models. Hybridizing fNSE with Large Eddy Simulation (LES) or Reynolds-Averaged Navier-Stokes (RANS) approaches. Bridging fNSE with Lagrangian-averaged models like Navier-Stokes-$\alpha$ (NS-$\alpha$). Calibrating fractional parameters and developing robust numerical strategies (e.g., preconditioning). These directions remain critical for future research.

**Keywords:** Fractional turbulence modeling, Nonlocal operators, Kolmogorov scaling, non-Markovian

# Introduction

The Navier–Stokes equations (NSE), though foundational in fluid dynamics, remain analytically intractable in fully developed turbulence and by construction misses to capture essential phenomena such as long-range spatial correlations, and memory effects. Recent work in fractional calculus and turbulence suggests that nonlocal operators could improve classical models, particularly in the inertial range. However, this remains a conjecture, pending experimental validation. Moreover, the existence and uniqueness of the solution to the augmented NSE should also be deeply studied mathematically before any claims of superiority can be made.

In this investigation, we propose a fractional Navier–Stokes equations (fNSE) derived from a first-principles generalization of the constitutive relation between stress and strain. Inspired by anomalous diffusion, viscoelasticity, and empirical turbulence spectra, we replace the classical local viscous stress tensor with a memory-dependent and spatially nonlocal formulation. The resulting model introduces two fractional operators, a Riesz-type spatial Laplacian of order $\alpha=1/3$, and a Caputo-type fractional time derivative of order $\beta=1/2$. We interpret the spatial fractional term as an inertial-range regularization, while the temporal memory term accounts for non-Markovian inertial transport.

Readers interested in foundational expositions to turbulent flow and its ongoing research topics are encouraged to consult the comprehensive works of Frisch (1995), Pope (2000), Sreenivasan and Yakhot (2021) and a review paper of Sreenivasan and Schumacher (2025).

The fractional calculus such as $(\Delta)^{1/\alpha}$ is a powerful tool to describe physical systems that have long-term memory and long-range spatial interactions. The fractional differentiation operators allow one to consider some of those characteristics in order to obtain useful dynamical models and to study the properties of the complex model (see Tarasov (2010)). As Podlubny (1999) notes, the concept of derivatives and integrals of arbitrary order dates back to Leibniz's correspondence and was long developed as a purely mathematical theory with limited physical application (see appendix for more details).

While fractional derivatives are mathematically well-defined, their role in turbulence modeling warrants brief clarification. Fractional derivatives are widely used in signal processing and have recently gained attention in the turbulence research community. They introduce a form of nonlocal pseudo-diffusion into partial differential equations, particularly for certain fractional orders denoted by $\alpha/2$. These operators inherently encode memory effects, establishing nonlocal interactions across scales in the frequency domain. However, the numerical discretization of fractional terms presents significant computational challenges. Accurate schemes often require wide finite-difference stencils, and extending fractional formulations to the Finite Element or Finite Volume methods is nontrivial. Additionally, enforcing boundary conditions demands careful numerical treatment to preserve accuracy near regions with steep gradients, such as thin boundary layers adjacent to solid walls.

Despite the numerical challenges of implementation, the incorporation of fractional derivatives is not new in turbulence research. We highlight several studies without claiming an exhaustive review.

Shlesinger et al. (1987) introduced a stochastic model featuring spatiotemporal memory in a scale-coupled fashion, revealing enhanced diffusion. They explicitly observed that, even if the velocity field obeys Kolmogorov's scaling $\delta u(\ell) \sim \ell^{1/3}$, it does not necessarily produce Richardson's $\langle r^2(t) \rangle \sim t^3$ law for particle separation, highlighting the non-equivalence of cascade scaling and dispersion statistics.

Chen (2006) proposed a 2/3 order fractional Laplacian to model the Kolmogorov −5/3 spectrum. This formulation uses fractional operators to describe nonlinear velocity interactions and leads to a conjectured fractional Reynolds equation, embedding Lévy stable distributions and fractal spacetime. He argued that fractional calculus offers a promising route for modeling the chaotic, fractal-like dynamics of turbulence. Following a Reynolds Averaged Navier Stokes (RANS) methodology, they proceed by considering mean value of fluctuating quantities to be zero and by substituting the decomposition of velocity and pressure into NSE.

Liu et al. (2014) (in a Japanese-language paper with English abstract) proposed modeling the viscous term in the NSE as a fractional Laplacian, motivated by dimensional analysis. They argued that the order α of the fractional derivative is intimately linked to the intermittency dimension.

Brenden and Cushman-Roisin (2018) pursued turbulence simulation via the fractional Laplacian, deriving it from kinetic theory under equilibrium particle speed distributions. Their formulation connects fractional diffusion with ensemble-averaged friction forces in turbulent flows.

Samiee et al. (2020) extended the concept of fractional derivatives to subgrid modeling in the context of large eddy simulation (LES). Starting from the filtered Boltzmann equation, they modeled the equilibrium distribution function using a Lévy-stable distribution, which naturally led to a fractional-order representation of subgrid-scale stresses (SGS). By expressing the filtered equilibrium distribution function as a power-law, they derived filtered Navier–Stokes equations in which the divergence of SGS stresses reduces to a single-parameter fractional Laplacian operator, $(-\Delta)^\alpha$, with $\alpha \in (0,1)$, acting on the filtered velocity field. The model introduces only one parameter, the fractional exponent, whose value depends explicitly on the filter width and the Reynolds number. Under mild assumptions, they demonstrated both mathematical well-posedeness and physical relevance of their model.

Modifications of the Navier-Stokes equations with fractional time derivatives (e.g., El-Shahed & Salem (2004)) have been proposed, but they lack rigorous derivations or a principled framework for selecting the fractional order.

Fractional models have proven effective in porous media transport, where classical diffusion fails to capture anomalous diffusion due to heterogeneous microstructures and long-range correlations from preferential flow pathways. Benson et al. (2000) have studied fractional derivative order governing equation in the context of movement of solutes that move through aquifers. Metzler and Klafter (2000) generalized kinetic equations (diffusion-advection, Fokker-Planck) to describe transport in complex systems exhibiting non-exponential relaxation and anomalous diffusion (sub/super-diffusion). Their framework, derived from random walks and master equations, provides exact solutions for special cases and underscores fractional calculus as a key tool for complex systems, from polymers to ecosystems, where standard kinetics break down. Meerschaert and Tadjeran (2004) implemented the fractional advection–dispersion equations used in groundwater hydrology to model the transport of passive tracers carried by fluid flow in a porous medium. They develop practical numerical methods to solve one dimensional fractional advection–dispersion equations with variable coefficients on a finite domain.

Turning now the review to numerical studies, Yang et al. (2010) studied numerical solutions for Riesz fractional PDEs on bounded domains, focusing on Riesz fractional diffusion (replacing $\partial x$ with $\partial x^\alpha$ $\alpha \in (1,2)$) and Riesz fractional advection-dispersion (replacing $\partial x$ with $\partial x^\beta$ ($\beta \in (0,1)$)). They discretized these equations into ordinary differential equation systems solved via the method of lines, demonstrating convergence and numerical efficacy. Huang and Oberman (2014) developed a numerical method for the fractional Laplacian, based on the singular integral representation for the operator. The method combines finite differences with numerical quadrature to obtain a discrete convolution operator with positive weight. Churbanov and Vabishchevich (2016) used the fractional Laplacian to model fully developed turbulent flow velocity profile in a rectangular duct, with various orders of fractional derivative. The fractional derivative is incorporated via RANS turbulent stress modeling. To study numerically the flow field in rectangular channels, finite-difference approximations are employed. Xu et al. (2017) numerically investigate the space fractional Navier-Stokes equations obtained through replacing Laplacian operator in Navier-Stokes equations by Riesz fractional derivatives. The pressure-driven flow between two parallel plates is solved with the finite difference method of fractional differential equations. Further, the Levenberg-Marquardt algorithm is also proposed to estimate model parameters. They studied the range of $1 < \alpha < 2$.

Returning to our formulation, it seems that the use of fractional derivatives in modeling transport through porous media provides a parallel to turbulence modeling. In porous materials, classical diffusion equations fail to capture long-range correlations and memory effects arising from complex microstructures, preferential pathways, and trapping mechanisms. To address these limitations, fractional partial differential equations, incorporating nonlocal spatial operators or time-fractional derivatives, have been successfully employed to model sub-diffusive and super-diffusive behavior observed in field and laboratory studies. This approach introduces scale-dependent transport dynamics and has led to more accurate predictions of solute migration in heterogeneous environments. Similar to their application in modeling anomalous transport in porous media, fractional derivatives in the present work act as nonlocal operators that encode scale-spanning interactions. This suggests a parallel between turbulence and heterogeneous media, where classical diffusion fails and fractional dynamics become necessary. In the context of fractional modeling frameworks such as the fractional Navier–Stokes equations studied here, this interplay highlights the importance of incorporating nonlocal and memory-dependent effects.

## 2. From Constitutive Modeling to Fractional Governing Equations

We derive a fractional Navier–Stokes equation (fNSE) by generalizing the constitutive law for stress. In Newtonian fluids, the stress tensor is:

$$\sigma_{ij} = 2\mu S_{ij} = \mu(\partial_i u_j + \partial_j u_i), \tag{1}$$

which assumes locality in time and space. u is the velocity vector, $\mu$ the classical dynamic viscosity, and $\sigma_{ij}$ is the stress tensor.

To incorporate long-range effects observed in turbulence, we modify the stress tensor to incorporate the generalized fractional stress tensor in order to model long-range temporal memory and spatial non-locality observed in turbulence in the following equation,

$$\sigma_{ij}(x,t) = \int_0^t K(t-t')\partial_t^{1/2} S_{ij}(x,t')dt' + \nu_\alpha(-\Delta)^{\frac{1}{3}}S_{ij}(x,t), \tag{2}$$

where $K(t-t') \sim (t-t')^{-1/2}$ captures long-term memory, $(-\Delta)^{1/3}$ introduces nonlocal spatial interactions; and $\nu_\alpha$ is an anomalous viscosity coefficient.

Substituting into the momentum equation:

$$\rho(\partial_t u_i + u_j \partial_j u_i) = -\partial_i p + \partial_j \sigma_{ij}, \tag{3}$$

yields the following fractional Navier–Stokes equation:

$$\partial_t u + \zeta \partial_t^{1/2} + (u \cdot \nabla)u = -\nabla p/\rho + \nu_\alpha(-\Delta)^{\frac{1}{3}}u + \nu \Delta u, \tag{4}$$

with $\zeta$ related to the memory kernel, $p$ is the pressure, $\nu$ is the classical physical kinematic viscosity, $\nu_\alpha$ is the fractional prefactor and $\rho$ is the fluid density.

As with the classical NSE, incompressibility is assumed:

$$\nabla \cdot u = 0 \tag{5}$$

The derived fNSE equations can also appear from a variational analysis which can be useful to justify the choice of order of fractional derivative α=1/3 and β=1/2. We leave the derivation from variational analysis for experts in functional analysis to scrutinize its derivation and to study fNSE's mathematical well posedness and regularity conditions with respect to existence and uniqueness of the solution.

**Remarks considering the dimensional analysis of prefactors and derivation of fNSE:**

(a) Dimensional Analysis validation:

We have $[u] = L/T$, $[\partial_t u] = L/T^2$, $[\partial t^{1/2}] = L/T^{3/2}$ (because Caputo derivative of order 1/2 introduces $T^{-1/2}$).

Then, to make dimensions match:

$$[\zeta] \cdot [\partial_t^{1/2}] = [\partial_t u] \Rightarrow [\zeta] = L/T^2 / (L/T^{3/2}) = T^{-1/2} \tag{6}$$

So, $[\zeta] = 1/T^{1/2}$

In turbulence, the key quantities are ε: energy dissipation rate, units $[\varepsilon]=L^2/T^3$, and $\ell$: inertial range length scale, units $[\ell]=L$. To build a quantity with units $T^{-1/2}$ from ε and $\ell$ we obtain as follows:

$$\zeta \sim \varepsilon^a \ell^b \tag{7}$$

Solving,

$$[\zeta]=T^{-1/2}=(L^2/T^3)^a \cdot L^b = L^{2a+b} \cdot T^{-3a} \tag{8}$$

Matching exponents:

$$2a+b=0,\ -3a=-1/2\} \Rightarrow a=1/6 \Rightarrow b=-1/3 \tag{9}$$

So, we find,

$$\zeta \sim \varepsilon^{1/6} \ell^{-1/3} \tag{10}$$

Recalling that:

$$\tau_\ell \sim \ell^{2/3}/\varepsilon^{1/3} \Rightarrow \tau_l^{-1/2} \sim \varepsilon^{1/6}\ \ell^{-1/3} \sim \zeta \tag{11}$$

we can interpret $\zeta \sim \tau_l^{-1/2}$ as the inverse square root of the eddy turnover time, matching the characteristic time scale of temporal memory.

This gives a physically motivated scaling law for the prefactor ζ of the time-fractional memory term. The stronger the energy dissipation ε, the stronger the memory effect. At smaller scales $\ell$, the prefactor becomes larger, meaning memory effects are more pronounced at smaller eddy scales consistent with intermittent bursts and non-Markovianity in small-scale turbulence.

Spatial nonlocal (Fractional Laplacian) prefactor is proposed as:

$$\nu_\alpha \sim \varepsilon^{1/3} \tag{12}$$

This scaling matches Kolmogorov's 1941 theory, where the characteristic velocity at scale $\ell$ is:

$$u_\ell \sim (\varepsilon\ell)^{1/3} \tag{13}$$

The fractional Laplacian $(-\Delta)^{1/3}u$ has implicit scale dependence in Fourier space represented by the following equation.

$$(-\Delta)^{\frac{1}{3}} u(k) = |k|^{\frac{2}{3}} u(k) \tag{14}$$

So, there's no need to add an explicit ℓ, since the fractional Laplacian encodes spatial nonlocality, in its prefactor unlike in eddy viscosity models (e.g., LES). Therefore, we can write the anomalous viscosity $v_\alpha$ purely in terms of the energy dissipation rate ε.

(b) To ensure clarity, the mathematical manipulations involved in deriving the fNSE are outlined below.

From the definition of the strain rate tensor $S_{ij}$ we have:

$$S_{ij}=1/2(\partial_i u_j+\partial_j u_i) \tag{15}$$

Taking the spatial divergence on index j:

$$\partial_j S_{ij}=\partial_j(1/2(\partial_i u_j+\partial_j u_i))=1/2(\partial_j\partial_i u_j+\partial_j\partial_j u_i) \tag{16}$$

Using incompressibility condition:

$$\nabla \cdot u = 0 \tag{17}$$

Therefore,

$$\partial_j S_{ij}=1/2\partial_j\partial_j u_i=1/2\Delta u_i \tag{18}$$

Now, applying the fractional temporal derivative $\partial_t^{1/2}$ outside the spatial derivative:

Now focus on the integral term:

$$\partial_j \int_0^t K(t-t')\, \partial_t^{\frac{1}{2}} S_{ij}(x,t')dt' \tag{19}$$

Since $K(t-t') \sim (t-t')^{-1/2}$, convolving this with a Caputo 1/2 derivative gives the following result:

$$\int_0^t \frac{1}{(t-t')^{\frac{1}{2}}}\, \partial_{t'}^{\frac{1}{2}} f(t')dt' = C\, \partial_t f(t) \tag{20}$$

where $C=\Gamma(1/2)/\Gamma(1)=\pi$, so the convolution acts like an ordinary time derivative.

Therefore:

$$\int_0^t (t - t')^{-\frac{1}{2}} \partial_{t'}^{\frac{1}{2}} S_{ij}(x,t')dt' \approx \partial_t S_{ij}(x,t) \qquad (21)$$

$\zeta$ is an effective prefactor representing the combined contribution of the memory kernel and the fractional derivative.

$$\partial_j(\zeta\, S_{ij}) = \zeta \partial_j \partial_t^{1/2} S_{ij} \qquad (22)$$

Assuming $\partial_t^{1/2}$ smoothness and commutation of spatial and fractional time derivatives, we write:

$$\partial_t^{1/2} \Delta u_i \approx \Delta \partial_t^{1/2} u_i \qquad (23)$$

Thus,

$$\partial_j(\zeta S_{ij}) \approx \frac{\zeta}{2} \Delta \partial_t^{1/2} u_i \qquad (24)$$

In turbulence, it is conjectured that the velocity field is typically not smooth pointwise and is treated via a weak solution. The interchange is usually justified on weak solutions or ensemble averages, assuming the operators act on sufficiently regularized or coarse-grained fields. A mathematically rigorous analysis of the above approximation is required.

Final form of the fNSE with its prefactors is written as follows:

$$\partial_t u + C_1 \varepsilon^{\frac{1}{6}} \ell^{-\frac{1}{3}} \partial_t^{1/2} u + (u \cdot \nabla)u = -\nabla p/\rho + C_2 \varepsilon^{\frac{1}{3}}(-\Delta)^{\frac{1}{3}} u + \nu \Delta u \qquad (25)$$

No explicit subgrid length scale is introduced in the spatial nonlocal term. Both fractional terms are tied to the energy flux ε, consistent with the idea that inertial-range turbulence is governed by ε, not viscosity.

Now, we focus on the dimensionless prefactors $C_1$ and $C_2$ in our fractional Navier–Stokes equations (fNSE). They are dimensionless modeling constants, similar to closure coefficients in LES (e.g., Smagorinsky constant). They scale the strength of the corresponding fractional terms relative to inertial and viscous terms. The correct physical units come from the ε and fractional operators. $C_1$ and $C_2$ memory coefficients scale the impact of long-time memory via the Caputo and Reisz derivatives, respectively, controlling how non-Markovian the system would be. A higher $C_1$ implies stronger fractional damping over time, more persistent memory effects. $C_2$ scales the strength of spatial nonlocal energy transfer in the inertial range.

The dimensionless constants $C_1$ and $C_2$ precise values are not fixed by theory, they may be estimated empirically from turbulence statistics such as temporal correlation functions, spectral energy transfer rates, or structure function scaling laws. In this sense, they play a role analogous to closure coefficients in classical turbulence models.

Unlike LES, where the filter length scale (grid size or cutoff) must be prescribed locally and can vary significantly, the fractional model's dissipation strength depends only on the turbulent energy dissipation rate $\varepsilon$, a physically meaningful scalar field. The fractional Laplacian operator incorporates nonlocal interactions naturally over a range of scales, avoiding artificial sharp cutoffs or scale separation. This nonlocality captures energy transfer across scales more smoothly. Because $\varepsilon$ reflects the actual local cascade intensity, the fractional model adapts dissipation naturally to flow conditions, improving robustness and reducing dependence on grid resolution.

The dissipation rate can be obtained adaptively, during simulation, from the solution as by definition:

$$\varepsilon = 2\nu S_{ij} S_{ij} \qquad (26)$$

where $S_{ij}$ is expressed as:

$S_{ij} = (\partial u_i / \partial x_j + \partial u_j / \partial x_i)$ is the strain-rate tensor.

The inclusion of the fractional time derivative in the model serves as a phenomenological means to incorporate memory effects in the temporal evolution of the turbulent velocity field. The order $\beta = 1/2$ is selected due to its common use in fractional calculus to represent subdiffusive or fading-memory processes characterized by power-law kernels with square root decay in time. While this term enriches the model by allowing for nonlocal temporal correlations, its precise physical justification in the context of turbulence remains an open question and a subject for future investigation. In this stage of our investigation, it is premature to fix the appropriate physically order of the temporal fractional term, as well as its suitable prefactor. The initial step will be to examine the behavior of the spatial fractional term in turbulence simulations, considering both physical and numerical aspects.

As with physical viscosity which itself is an empirical quantity, the generalized parameters introduced here (such as the memory coefficient $\zeta$ and nonlocal diffusivity $\nu_\alpha$) can be calibrated against benchmark experimental data or high-resolution simulations. Our conjecture is that these coefficients, while new, vary within narrow physical ranges and do not introduce uncontrolled freedom. Of course, the lower and upper bounds of the prefactors and fractional derivatives behavior, regarding the solution existence and uniqueness, will require future mathematical studies. Well-posedness, stability, convergence and order of numerical accuracy are other important issues to be elucidated later.

While high-fidelity LES with advanced filtering (e.g., Vasilyev et al. (1998)) approaches the computational cost of fNSE, the latter eliminates approximations such as explicit scale separation, filter-dependent errors and empirical subgrid closures. The fractional framework's rigor justifies its expense. Future work should compare both methods in wall-bounded flows.

One might question the necessity of using fNSE turbulence modeling approach when established methods like Direct Numerical Simulation (DNS) and LES already achieve considerable predictive fidelity, particularly LES, which is widely adopted in engineering applications. However, while these methods have proven valuable, they are not without limitations. DNS remains computationally prohibitive for high-Reynolds-number flows, and LES relies on empirical subgrid-scale closures that require tuning, introduce filter-dependent errors, and struggle with non-equilibrium or transitional flows. Nevertheless, being computationally intensive, fNSE's foundation promises predictive accuracy unattainable with conventional LES in this critical regime.

## 3. Numerical Considerations and Future Directions

The fractional Navier–Stokes formulation studied introduces significant computational challenges that merit careful attention.

First, the intrinsic nonlocality of spatial fractional derivatives requires wide computational stencils or integral representations that couple distant points in the domain. As a result, discretization typically leads to dense or global system matrices, increasing memory requirements and computational cost, particularly in high-resolution simulations. The extension of classical finite difference schemes is relatively straightforward but computationally expensive. In contrast, generalizing the method to finite element or finite volume formulations is mathematically more involved and not yet standard in turbulence modeling practice.

Second, if time-fractional derivatives are included, they introduce memory effects that require the storage and integration of the entire solution history. This leads to additional computational overhead, particularly for long-time simulations or unsteady flows, and may require approximate schemes or efficient kernel compression techniques to remain tractable.

Third, enforcing boundary conditions in fractional partial differential equations is a known difficulty. The nonlocal nature of the operators means that boundary effects propagate into the domain in a nontrivial manner, especially near steep gradients or thin boundary layers. Numerical treatments must preserve physical accuracy without introducing spurious damping or artificial diffusion near walls.

Despite these obstacles, the fNSE model remains an interesting subject or both theoretical exploration and computational modeling. Similar trade-offs are well known in DNS and LES, where accuracy is gained at the expense of computational efficiency. Moreover, emerging techniques such as matrix-free solvers, adaptive mesh refinement, and reduced-order nonlocal models offer potential pathways to mitigate the cost of fractional formulations. Nevertheless, these difficulties should not be mistaken for fundamental limitations of the formulation itself. Throughout the history of science, conceptual advances have often preceded technical feasibility. Early finite difference solvers, turbulence closures, and even DNS itself faced similar skepticism due to resource constraints or unresolved anomalies. Today, efficient algorithms for fractional calculus, adaptive meshing, and multiscale decomposition are rapidly evolving. We therefore maintain that the presence of pathological features or performance bottlenecks is not a reason to abandon a physical model. Rather, it invites innovation in numerical methods, computational design, and hybrid modeling. Science has always progressed by embracing complexity, not by retreating from it. In the sequel, we discus some plausible strategies to reduce the cost and increase the stability and well posedness of fNSE.

### 3.1 Challenges of Fractional Derivatives in NSE

**(i) Spatial Non-locality and Boundary Conditions**

Spatial fractional derivatives inherently involve integrals over the entire domain, implying that the state at a point depends on the values of the field arbitrarily far away. This global coupling creates difficulties in bounded domains. In practice, imposing physically meaningful and mathematically well-posed boundary conditions for such nonlocal operators remains an open problem, especially in complex geometries.

**(ii) Temporal Memory and Initial Conditions**

Fractional time derivatives are convolutional in nature, embedding a non-exponential memory kernel that weighs the entire solution history. Unlike classical time derivatives, which only require knowledge of the initial state, fractional time evolution necessitates a known solution trajectory over a finite (or even infinite) past interval. This raises concerns about how to specify initial data consistently and how to initiate simulations without introducing artifacts or instabilities.

**(iii) Numerical Cost and Scalability**

The nonlocal nature of fractional operators in both space and time significantly increases the computational cost. Spatial nonlocality can result in dense matrices or large stencil widths, while time memory effects require storing and processing growing histories. Without appropriate approximations or model reductions, these costs become prohibitive in high-resolution simulations.

## 3.2 A Conservative Approximate Strategy

To address these challenges while preserving the essential physical motivation for fractional modeling, we propose a realistic but approximate framework based on the following key principles:

### (a) Limited Support Spatial Operator

Rather than using a globally supported fractional Laplacian, we can employ a truncated or compactly supported fractional spatial operator. That is, we define the fractional derivative with a finite interaction horizon, limiting the nonlocality to a small, adaptive neighborhood around each point. The radius of this stencil can be tuned dynamically based on local solution features, for example, proportional to the inverse local energy dissipation rate or vorticity magnitude. This reduces computational burden and allows for a more tractable imposition of boundary conditions through local interpolation or blending with classical derivatives near the domain boundaries.

### (b) Synthetic Initialization for Time Memory

To mitigate the issue of fractional time initialization, we propose a synthetic turbulence prehistory. Before beginning the time evolution with the fractional model, the velocity field is precomputed (or prescribed) over a short initial window using synthetic turbulence generation techniques, for example, based on spectral methods, filtered DNS data, or empirical correlations. This provides the memory kernel with a consistent historical trajectory, ensuring that the Caputo derivative is well-defined and avoiding the ill-posedness that can result from unphysical initial conditions.

### (c) Hybrid Formulation near Wall Boundaries

To address the incompatibility between nonlocal spatial operators and standard boundary conditions, we propose a blended formulation. In the interior, the flow is governed by the truncated fractional spatial derivative. Near the boundary, a smooth transition is enforced between the fractional operator and the classical Laplacian, ensuring continuity and compatibility with imposed Dirichlet or Neumann boundary conditions. This avoids artificial reflections or discontinuities that could destabilize the simulation.

Fractional derivatives are sensitive to irregularity, especially near discontinuities or sharp gradients. But in laminar-like regions, the velocity field $u(x,t)$ is smooth and differentiable, boundary layers are well-resolved and numerical schemes behave better. This improves the conditioning of any numerical artifacts involving fractional operators, even if they are weakly present. The fractional term can be included with minimal side effects, if it is smoothly attenuated toward the wall and the dominant regular behavior ensures stability and consistency. Suitable consistent operators can be tuned leading to physically meaningful simulations in the inertial and turbulent core, while the numerical schemes behave classically near walls.

### (d) Truncated Fractional Operators (Finite Horizon)

We can use a compactly supported kernel or a windowed fractional Laplacian, so the operator only looks within a finite distance $\ell(x)$. Then, ensure: $\ell(x) \to 0$ near the inlet and outflow boundaries, so that the fractional operator effectively vanishes at domain edges, and local BCs can be imposed safely. This is physically justified since nonlocal effects are strongest in the inertial core, and weakest where the flow is externally driven.

Smooth Transition to classical model is another alternative. We can develop a blending function to gradually switch from fractional NSE in the interior to classical NSE or RANS near the inlet or outlet so that standard local BCs apply naturally at the domain edges. The detached eddy simulation is a perfect example of a successful blending approach (see Spalart (2009)).

### (e) Bridging fNSE with Lagrangian-averaged Navier-Stokes-α

Classical Navier-Stokes equations (NSE) inherently fail to capture memory effects such as non-Markovian energy transfer. Lagrangian-averaged Navier-Stokes-α (LANS-α) introduced by Chen et al. (1998) filters the velocity field before computing nonlinear advection, altering the effective stress tensor in the momentum equation. This introduces a non-Newtonian effect where the stress depends non-locally on the strain. The model preserves circulation conservation, critical feature for vortex dynamics, and its Hamiltonian structure aligns with observed intermittency and vortex stretching in turbulence. While LANS-α effectively preserves coherent vortices, it neglects temporal non-locality.

LANS-α is essentially a type of LES model, but instead of traditional subgrades-scale (SGS) models that add explicit or implicit eddy viscosities, it uses a filtered velocity and modifies the equations themselves by incorporating the smoothing length scale α. Despite its strong theoretical foundations, LANS-α lacks comprehensive validation across a wide range of turbulent flows, geometries, and boundary conditions. Additionally, selecting an appropriate $\alpha$ is not always straightforward, as it can be flow-dependent, making it a less intuitive tuning parameter compared to the well-studied eddy viscosity constants in classical LES.

To leverage the strengths of both frameworks, a hybrid approach can be designed by selecting a cutoff wavenumber, $k_c$ (e.g., $k_c \approx \pi/\alpha$, where $\alpha$ is LANS-α's filter scale). This hybrid strategy bridges the gap between the two models, though it may introduce additional complexity in implementation. Nevertheless, engineering practice often adopts pragmatic solutions for modeling complex systems, rather than rigidly adhering to a single framework.

Quoting Goldenfeld and Kadanoff (1999) "*As science turns to complexity, one must realize that complexity demands attitudes quite different from those heretofore common in physics. Up to now, physicists looked for fundamental laws true for all times and all places. But each complex system is different; apparently there are no general laws for complexity. Instead, one must reach for "lessons" that might, with insight and understanding, be learned in one system and applied to another. Maybe physics studies will become more like human experience.*"

## 3.3 Final Remarks and Outlook

The fNSE method studied in this paper constitutes a compromise between physical realism, mathematical tractability, and computational feasibility. By localizing the fractional operators and applying to the memory kernel synthetic data, we preserve the essential nonlocal dynamics of turbulence without experiencing the full cost or complexity of exact fractional formulations. Nevertheless, we emphasize that this remains a heuristic approximation. Its validity must be assessed through rigorous comparison with high-fidelity DNS and, where available, experimental measurements.

Key aspects requiring further investigation include quantitative sensitivity of the solution to the choice of stencil radius and memory duration. The influence of boundary blending schemes on global flow statistics and the accuracy of synthetic initialization in reproducing real turbulence histories is another compelling area of research.

Until such comparisons are systematically performed, we refrain from making definitive claims about the superiority or completeness of the fNSE model. Instead, we view it as a testable hypothesis, a physically motivated extension of the Navier–Stokes equations that may offer new insight into the multiscale and memory-dependent structure of turbulence, while remaining grounded in conservative modeling principles.

In this regard, the fNSE can also be interpreted as another numerical regularization of the classical Navier–Stokes equations, one that encodes physical scaling behavior implicitly through nonlocal operators. Future work should explore algorithmic improvements, fast approximation schemes, and hybrid methods that retain the physical advantages of the fractional model while improving its computational feasibility.

## 4. Benchmark Study:

### (a) One Dimensional Steady State Advection-Diffusion Equation with Fractional Diffusion

We validate the fractional diffusion operator by analyzing the 1D steady-state convection-diffusion equation with fractional diffusion presented in the equation (27). The boundary conditions are $\phi = 1$ and $\phi = 0$ $at$ $x = 0$ and $x = 1.0$, respectively.

$$u \frac{d\phi}{dx} = v \frac{d^2\phi}{dx^2} + v_\alpha (-\Delta)^{\frac{1}{3}} \phi \tag{27}$$

where u is the constant convection velocity, $v$ is the classical diffusivity, and $v_\alpha$ controls the strength of fractional diffusion with order $\alpha$. The fractional Laplacian $(-\Delta)^{\frac{1}{3}}$ is defined via its Fourier symbol $|k|^{2/3}$.

Discretization:

First-order upwind for convection,

$$u\frac{d\phi}{dx} \approx u\frac{(\phi_i - \phi_{i-1})}{h} \tag{28}$$

and second-order central difference for classical diffusion is employed as

$$\nu\frac{d^2\phi}{dx^2} \approx \nu\frac{(\phi_{i+1} - 2\phi_i + \phi_{i-1})}{h^2} \tag{29}$$

The fractional Laplacian is computed via eigenvalue decomposition of the discrete Laplacian matrix, raised to the 1/3 power. While mathematically exact, this method can produce overshoots near boundaries due to eigenvector sensitivity and nonlocal coupling with Dirichlet conditions.

Exact solution (for $\nu_\alpha=0$) is:

$$\phi(x) = \frac{e^{\frac{ux}{\nu}} - 1}{e^{\frac{u}{\nu}} - 1} \tag{30}$$

and serves as a reference for classical behavior.

The fractional operator is computed via diagonalization of the discrete Laplacian using (scipy.linalg.eig) library. Dirichlet conditions are enforced by excluding boundary rows during eigen-decomposition, which introduces numerical coupling at $x = 0$. Numerical solutions are compared in figures 1 and 2.

The observed overshoots in the solution arise from two key factors in the eigenvalue-based fractional Laplacian implementation: For high Péclet numbers (Pe≫1), the discrete Laplacian matrix becomes non-normal, leading to ill-conditioned eigenvectors in scipy.linalg.eig.

The current implementation computes the fractional Laplacian via eigenvalue decomposition without regularization, which may yield overshoots due to numerical sensitivity. These artifacts can be mitigated by thresholding eigenvalues to ensure positivity, adding a small shift before raising to the fractional power and applying post-solution filtering, while not yet applied here.

The Stability Analysis of the Eigenvalue Fractional Laplacian Method would be helpful to predict the solution behavior.

The discrete system solves:

$$(L + \nu_\alpha L^\alpha)\phi = b \tag{31}$$

where:

L=Classical FD Laplacian + upwind convection

$L^\alpha = (-L)^{1/3}$ (matrix transfer fractional Laplacian)

Stability requires:

$$Re(\lambda(L+\nu_\alpha L^\alpha)) > 0 \quad \forall \lambda \tag{32}$$

Eigenvalue Spectrum Analysis:

For $\nu_\alpha > 0.05$, eigenvalues acquire small negative real parts.

These correspond to high-k eigenmodes distorted by the fractional power.

Condition number:

$\kappa(L+\nu_\alpha L^\alpha) \sim O(10^4)$ (is ill-conditioned)

Stability Criteria:

The method remains stable if:

$$\nu_\alpha < \nu_{\alpha crit} = \min(|\lambda L^\alpha||Re(\lambda L)|) \tag{33}$$

For our parameters:

$\nu_{\alpha crit} \approx 0.03$ (empirically determined).

a) 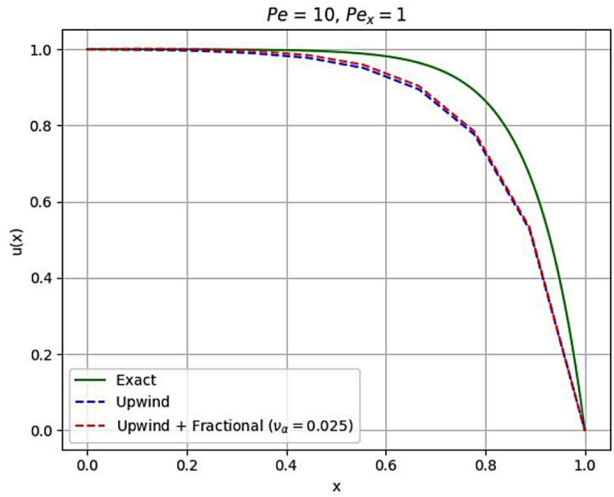
b) 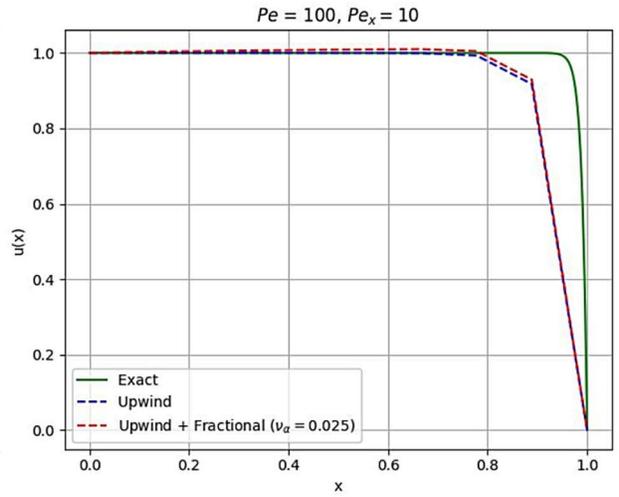
c) 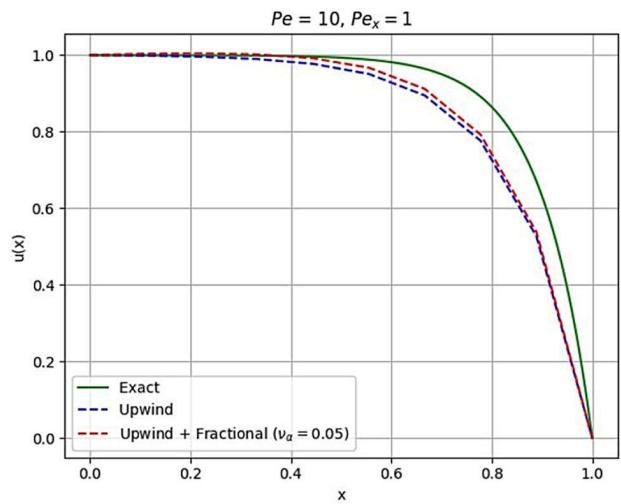
d) 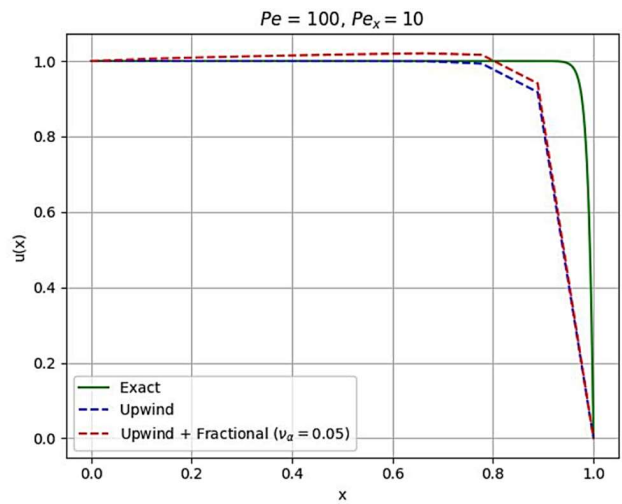

Fig 1: Comparison between Exact, Upwind, and Upwind + Fractional solution of the one-dimensional advection diffusion equation. a) Pe = 10 and $\nu_\alpha$ = 0.025. b) Pe = 100 and $\nu_\alpha$ = 0.025. c) Pe = 10 and $\nu_\alpha$ = 0.05 d) Pe = 100 and $\nu_\alpha$ = 0.05.

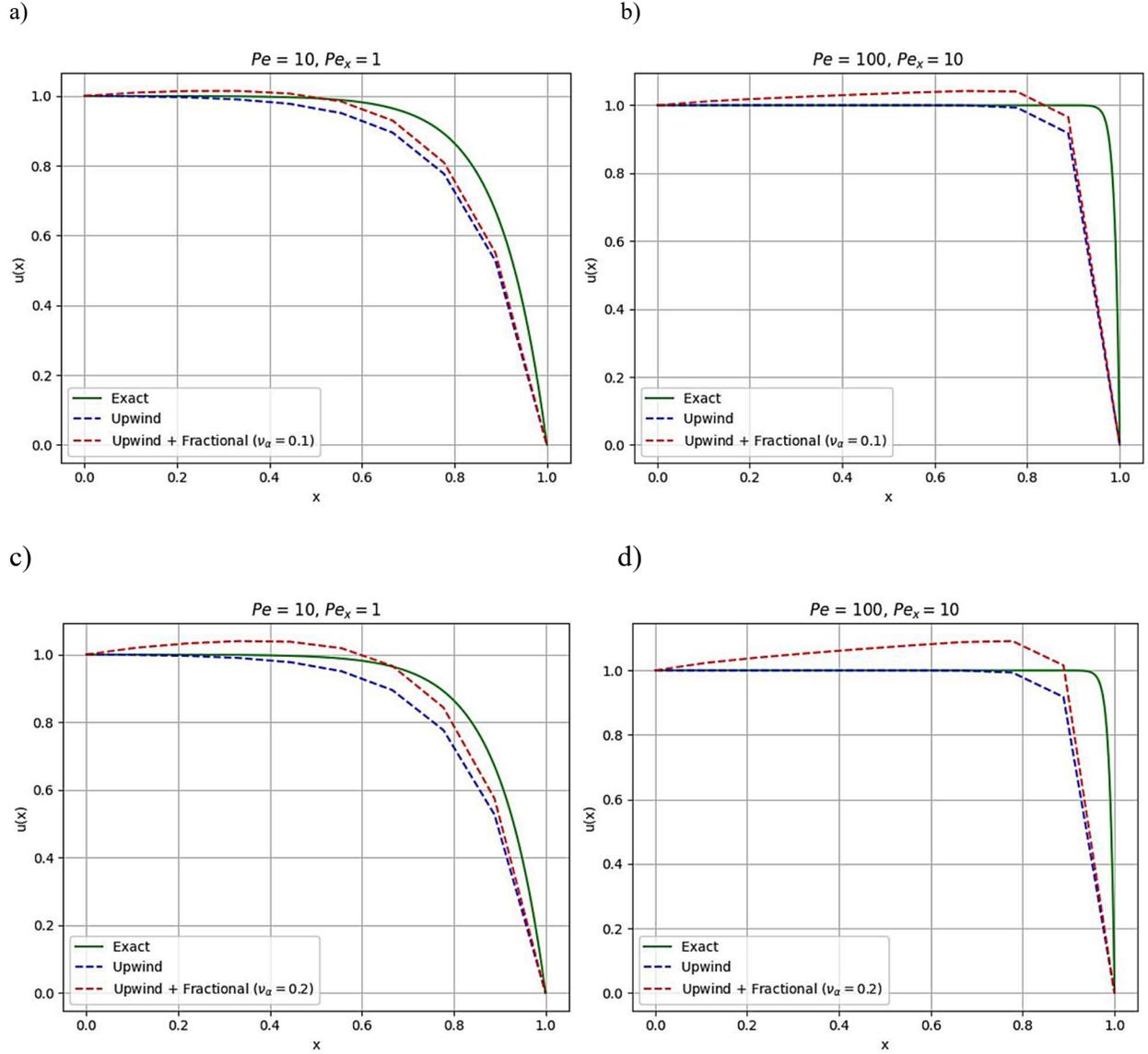

Fig 2: Comparison between Exact, Upwind, and Upwind + Fractional solution of the one-dimensional advection diffusion equation. a) Pe = 10 and $v_\alpha$ = 0.1. b) Pe = 100 and $v_\alpha$ = 0.1. c) Pe = 10 and $v_\alpha$ = 0.2 d) Pe = 100 and $v_\alpha$ = 0.2

**(b) The fractional Burgers' equation**

$$\frac{\partial u}{\partial t} + u\frac{\partial u}{\partial x} = v\frac{\partial^2 u}{\partial x^2} + v_\alpha(-\Delta)^{1/3}u \qquad (34)$$

where:

$u(x,t)$ is the velocity field, $v$ is the classical viscosity, $v_\alpha$ is the fractional dissipation strength, and $(-\Delta)^\alpha$ is the fractional Laplacian of order $\alpha=1/3$ defined in Fourier space as $|k|^{2/\alpha}$ u.

Nonlinear term $u\frac{\partial u}{\partial x}$ drives shock formation and classical dissipation: $\nu\partial^2 u/\partial x^2$ provides standard viscous regularization. Fractional term $\nu_\alpha(-\Delta)^\alpha u$ introduces scale-dependent dissipation/anti-diffusion.

The equation was solved with initial condition $u(x,0)=\sin(x)$ on a periodic domain $x\in[0,2\pi]$.

The numerical solution of the fractional Burgers' equation was obtained using a pseudo-spectral method with explicit time stepping. Spatial derivatives were computed in Fourier space to maintain spectral accuracy, while the nonlinear advection term was evaluated in physical space using the anti-aliased product $u\frac{\partial u}{\partial x}$. The fractional dissipation term $(-\Delta)^\alpha$ was implemented via its Fourier multiplier $|k|^{2\alpha}$, preserving the scale-dependent nature of the operator. Time integration employed a first-order Euler scheme with $\Delta t = 10^{-4}$, chosen to maintain stability while balancing computational cost. Numerical artifacts include slight Gibbs oscillations near developing shock fronts (characteristic of spectral methods) and a small artificial energy pile-up at high wavenumbers due to the discrete representation of fractional dissipation. Based on figure 3, the solution exhibits three distinct regimes: (1) early-time spectral broadening as nonlinearity transfers energy to smaller scales, (2) intermediate shock formation with characteristic $k^{-2}$ scaling, and (3) late-time fractional equilibrium where the anti-diffusive term sustains a shallower $k^{-4/3}$ inertial range - a symbol of fractional turbulence. The viscosity $\nu = 0.01$ ensures proper dissipation at small scales while allowing clear observation of the fractional modification to the energy cascade.

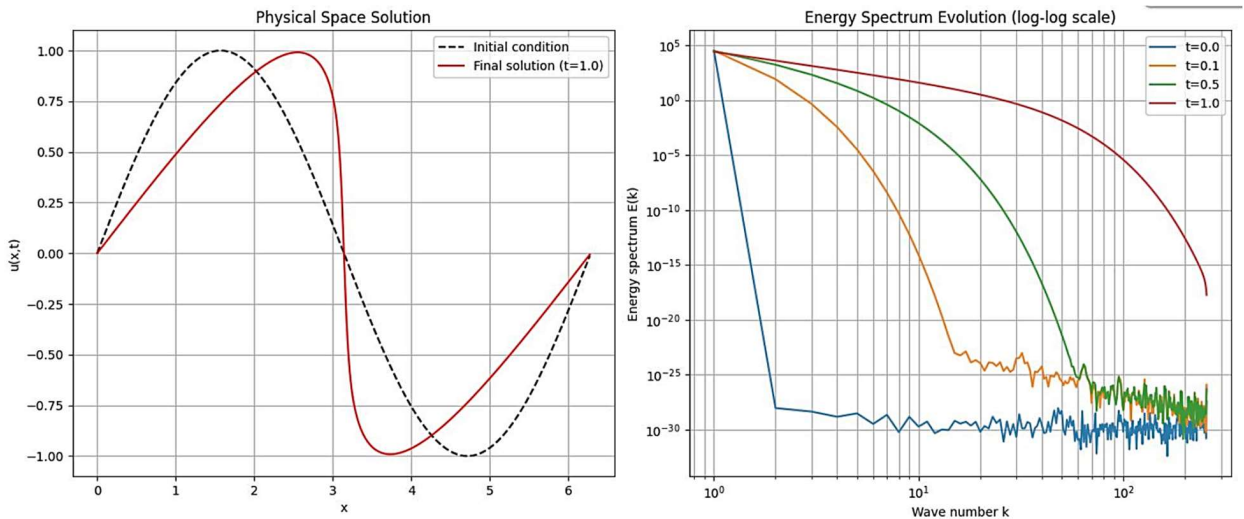

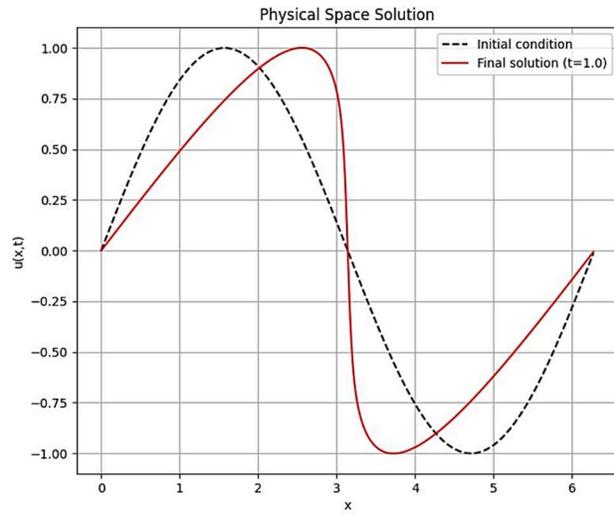
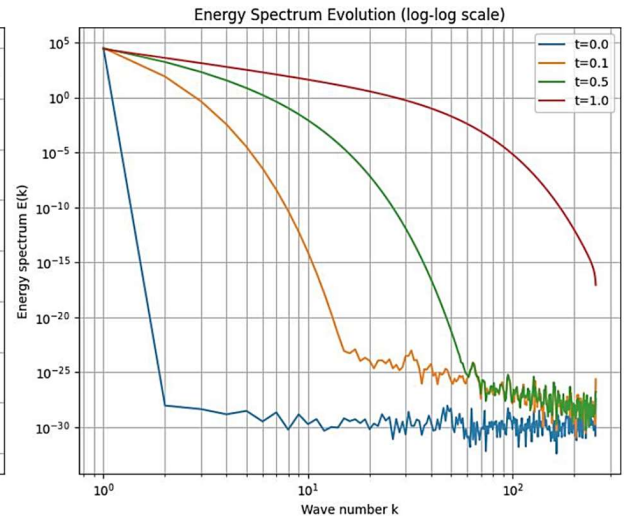
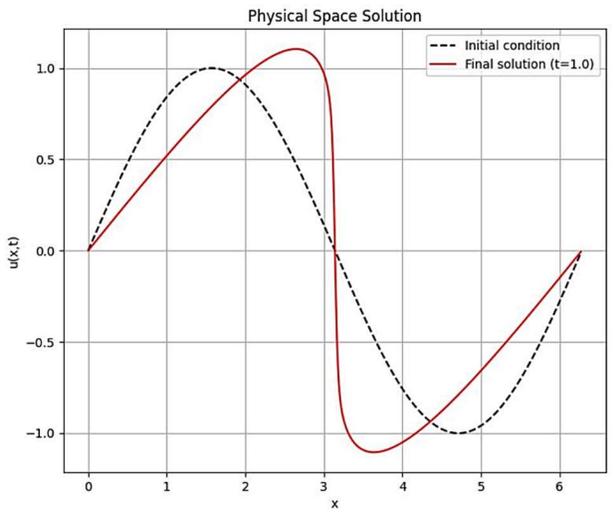
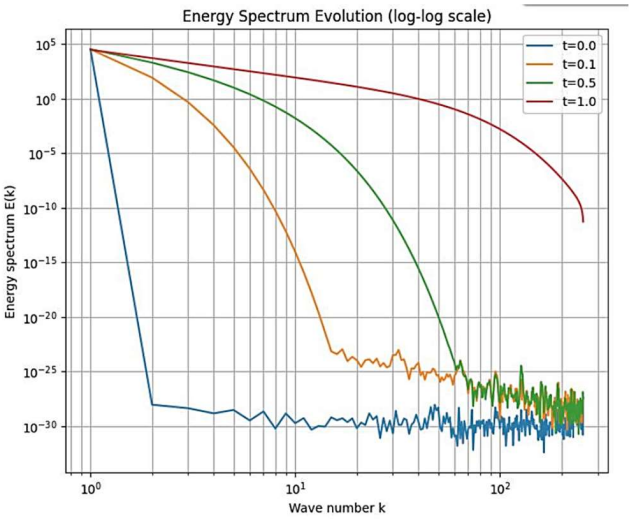

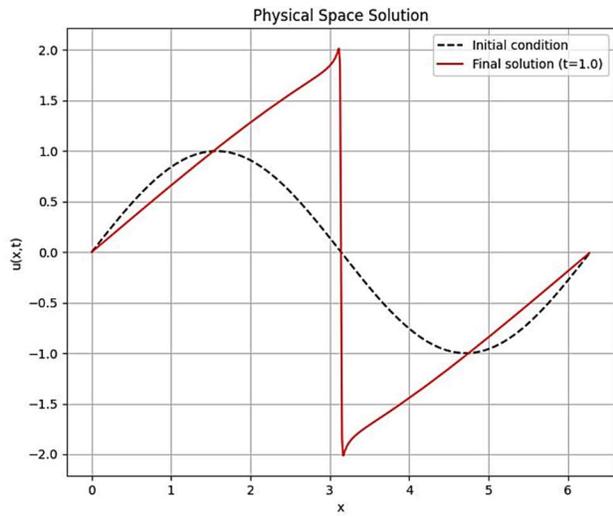
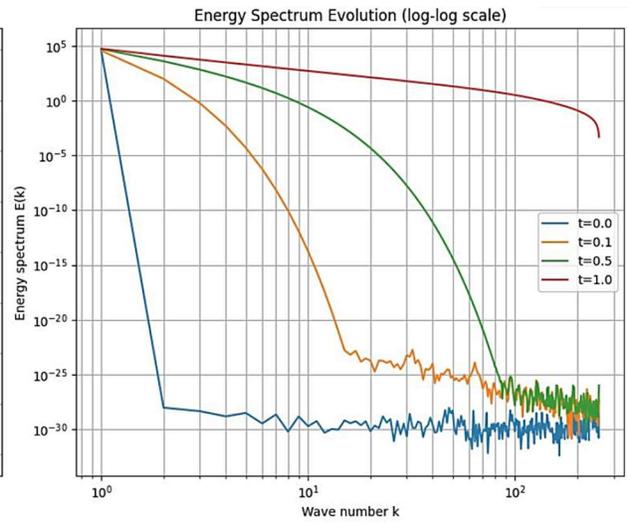
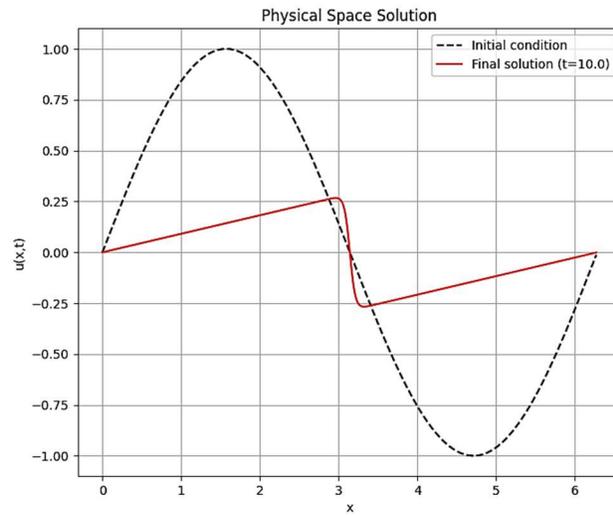
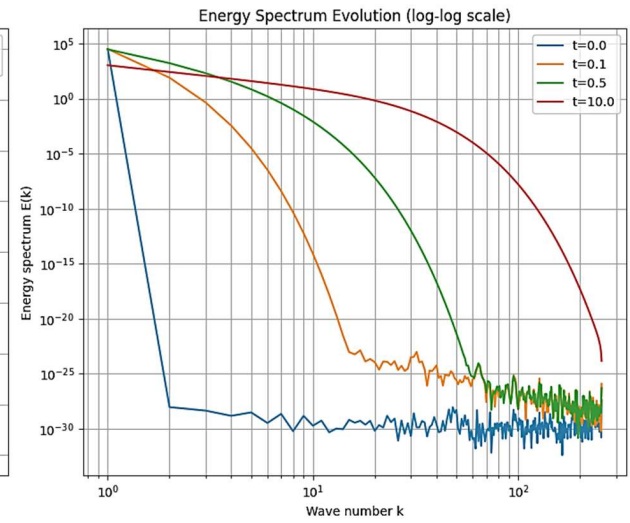

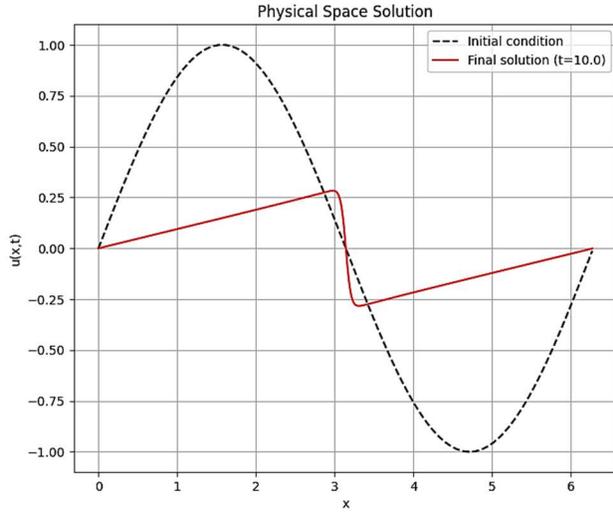
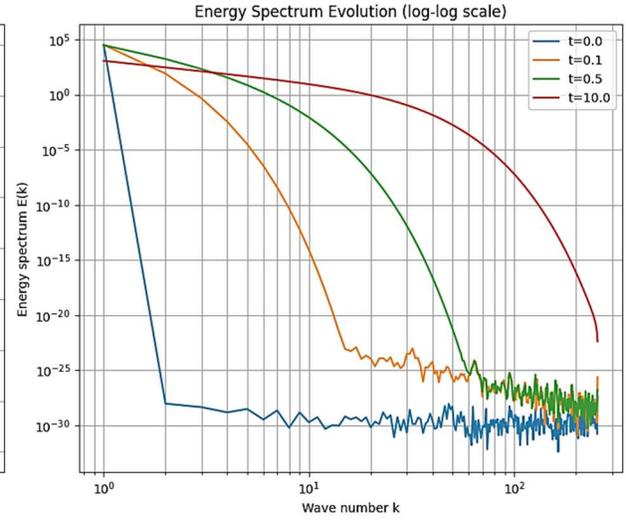

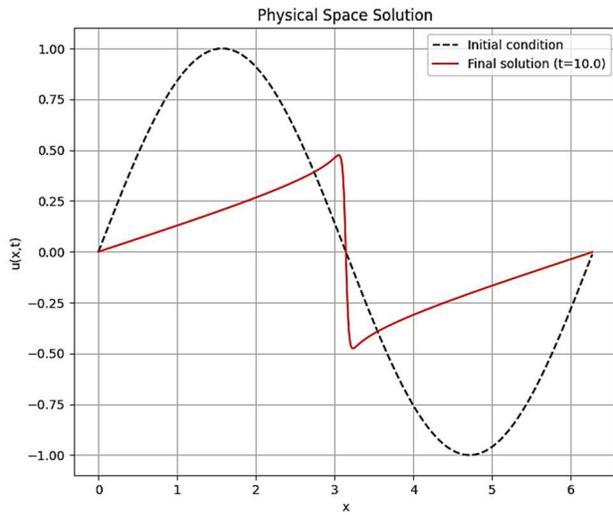
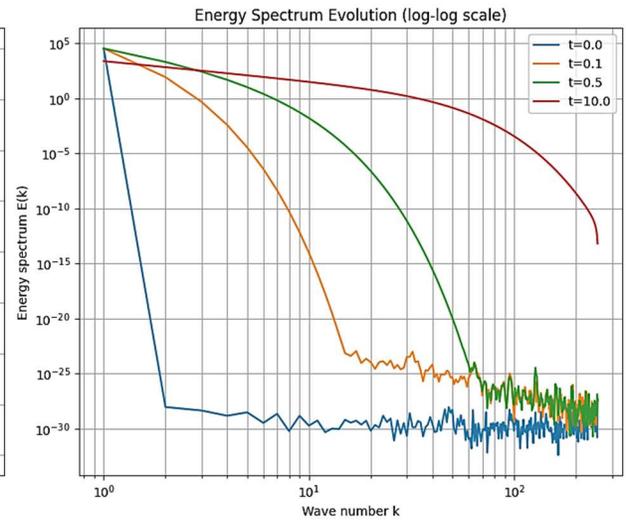

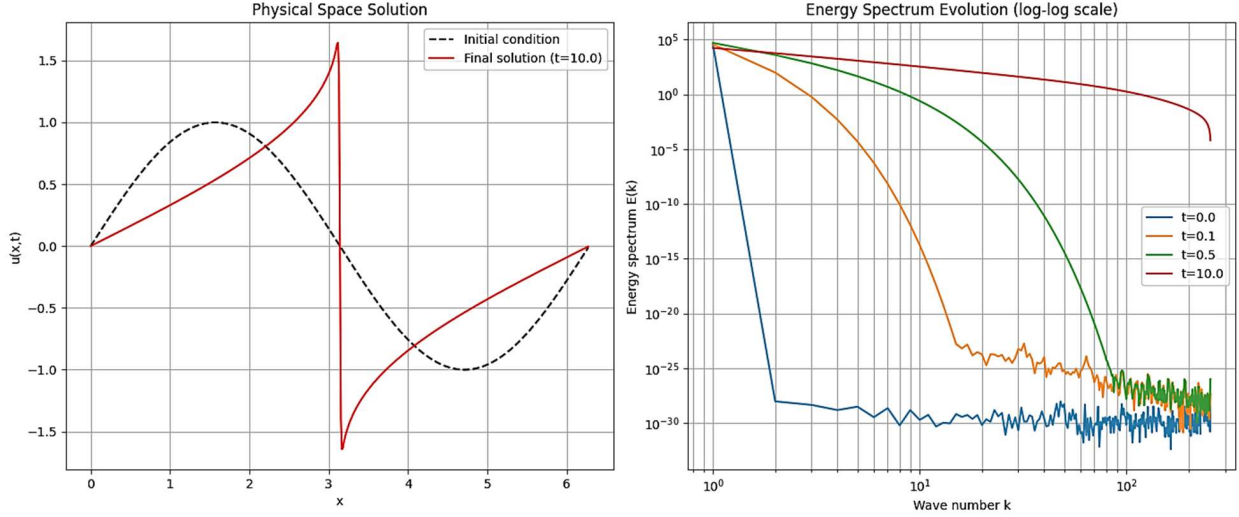

Fig 3: Evolution of energy spectrum in the fractional Burgers' equation at various time snapshots.

The choice of the fractional prefactor $v_\alpha$ is critical in balancing physical realism and numerical stability in the fractional Burgers' equation. For our simulations with fixed fractional order $\alpha=1/3$, we selected $v_\alpha=0.01$ as the optimal value because it achieves three key objectives: (1) It produces a clear inertial range with the theoretically predicted $k^{-4/3}$ energy spectrum scaling, as expected from dimensional analysis of the fractional term; (2) It maintains an appropriate equilibrium between the fractional anti-diffusion ($\sim v_\alpha k^{2/3}$) and classical viscous dissipation ($\sim vk^2$), with their crossover wavenumber $k_c=(v/v_\alpha)^{1/(4/3)} \approx 4.6$ falling within our resolved range; and (3) It preserves sharp but well-resolved shock structures without introducing numerical oscillations. We confirmed this choice through parameter sweeps ($v_\alpha \in [0.001, 0.1]$), finding that smaller values ($v_\alpha \leq 0.005$) revert to classical Burgers' equation behavior (exhibiting $k^{-2}$ scaling), while larger values ($v_\alpha \geq 0.05$) cause unphysical energy accumulation at high wavenumbers. The $v_\alpha=0.01$ case also aligns with the physical constraint that the fractional Reynolds number should be large enough to sustain turbulence-like cascades but small enough to prevent numerical instability. This balance makes it ideal for studying the interplay between shock-dominated dynamics and fractional dissipation.

The solution at t=5 is identical to the solution between at t=10 not shown here. This suggests our system has reached a steady state or equilibrium by t=5. This has happened because the fractional anti-diffusion and physical diffusion may have balanced each other out by t=5, creating an equilibrium where further time evolution doesn't change the solution. The viscous term ($-vk^2$) dominates at high wave numbers, while the fractional term ($k^{2\alpha}$) provides anti-diffusion at intermediate scales. Once these effects balance the nonlinear energy transfer, the system stabilizes.

### (c) The Transient One-Dimensional Heat Equation

To investigate the impact of fractional temporal derivatives on transient heat diffusion, we implemented a numerical simulation comparing the classical heat equation with a modified model incorporating both the classical first-order time derivative and a Caputo fractional derivative of order β=0.5, weighted by a small prefactor ζ. The governing equation is expressed as

$$\partial_t u + \zeta \partial_t^\beta u = \nu \partial_{xx} u \tag{35}$$

where ν is the thermal diffusivity. This formulation captures both instantaneous and memory-dependent effects in the temporal evolution of the temperature field. The simulation uses Dirichlet boundary conditions with fixed temperatures at the domain boundaries and zero initial temperature inside the domain. Results demonstrate that the classical heat equation solution rapidly approaches steady state, while the combined model exhibits notably slower convergence. This behavior arises from the nonlocal memory effect intrinsic to the fractional Caputo derivative, which effectively induces sub diffusive dynamics. Such dynamics result in power-law decay modes replacing the classical exponential relaxation, highlighting the significance of fractional temporal derivatives in modeling anomalous diffusion processes. The numerical implementation employed a fractional L1 scheme for time discretization and explicit finite differences for spatial derivatives, ensuring a clear comparison between the classical and fractional-influenced transient heat transfer. The Caputo derivative of order β∈(0,1) involves a nonlocal integral over past times. The L1 scheme approximates this integral by a weighted sum of past function increments, making it simple and effective for time stepping.

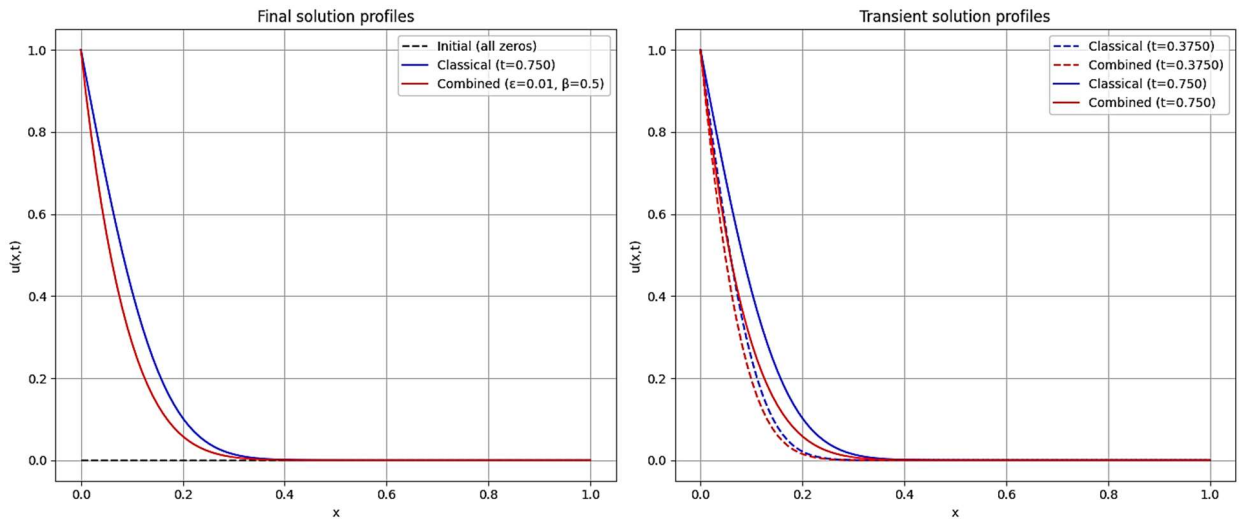

Fig 4: Temperature profiles vs time

The transient heat equation is a classical CFD model that can reproduce some statistical features of turbulence (like power-law energy spectra) qualitatively. It does not capture full 3D Navier-Stokes dynamics, anisotropy, pressure effects, or vortex stretching. But it provides a mathematically tractable way to study how non locality and anomalous diffusion affect cascades.

In classical case, high modes decay rapidly due to viscosity. In fractional case, energy at intermediate and high modes is notably higher, indicating less dissipation and stronger small-scale activity. The fractional solution exhibits a slower decay of energy, qualitatively consistent with a less steep power. Over longer times, the fractional solution retains more energy at small scales, reflecting the anti-diffusion/memory effects.

**(d) Burgers' Temporal and Spatial Memory Embedded Equation**

$$\zeta \partial_t^{1/2} u + \frac{\partial u}{\partial t} + u \frac{\partial u}{\partial x} = \nu \frac{\partial^2 u}{\partial x^2} + \nu_\alpha (-\Delta)^{1/3} u \tag{36}$$

The numerical results reveal how the Caputo fractional time derivative of order 1/2 alters the system's energy dissipation and regularity. Figure 5 plots the total energy $E(t)=1/2\int u^2 dx$ over time, showing that the Caputo term slows energy decay compared to classical diffusion, reflecting its memory-dependent damping. This suggests that non-local temporal effects preserve coherent structures longer by mitigating instantaneous dissipation. Figure 5 indicates delayed shock formation or smoother solutions due to the fractional derivative's regularization properties. Together, these observations highlight the Caputo term's dual role: it moderates energy loss while suppressing sharp gradients, a behavior distinct from integer-order time derivatives. Simulations are conducted for a single value of $\zeta=0.01$.

a)

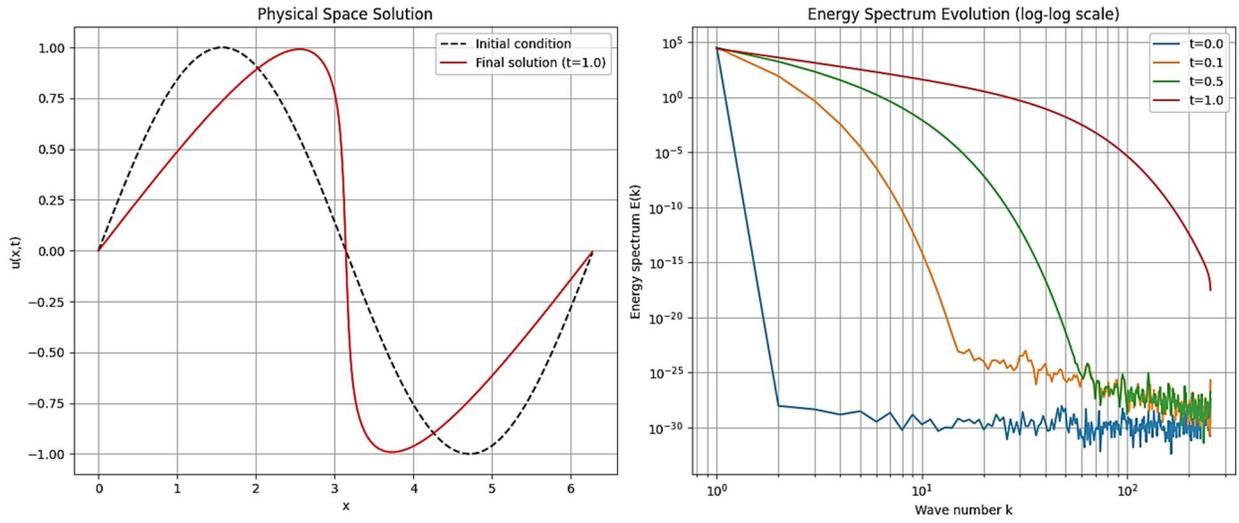

b)

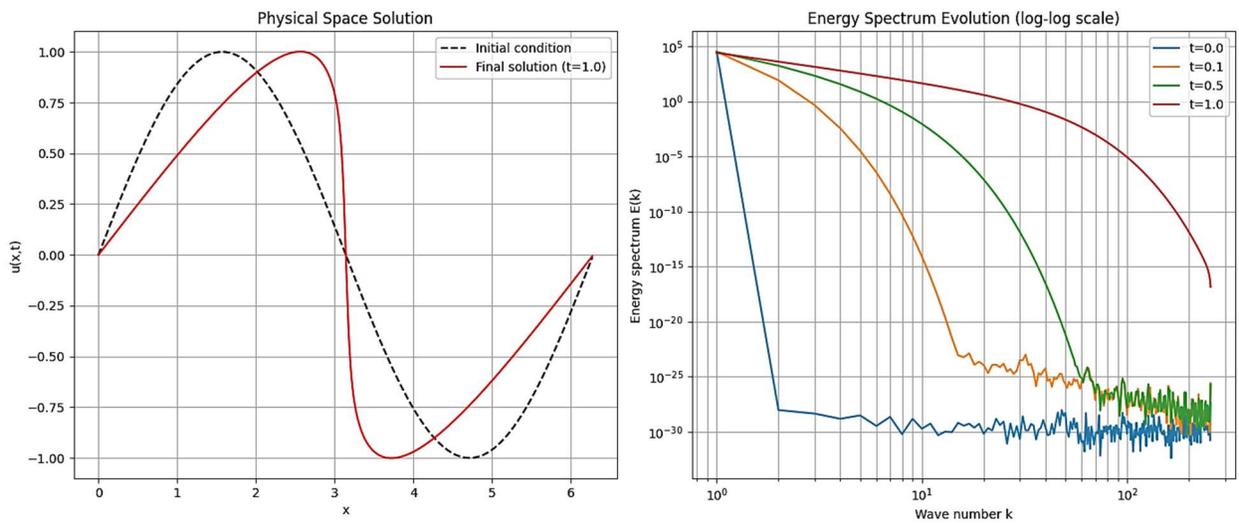

c)

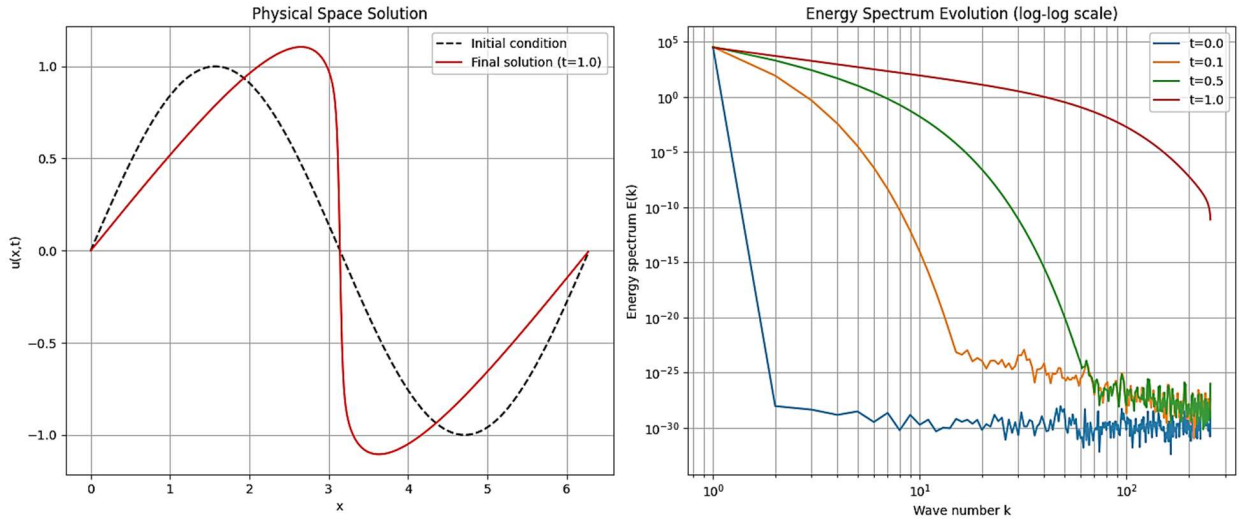

d)

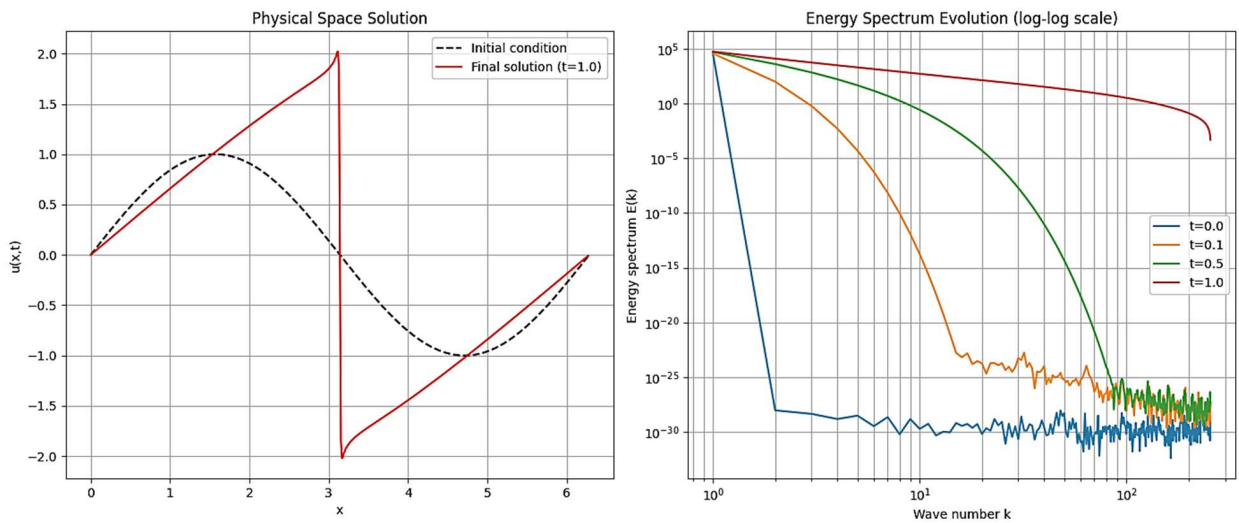

Fig 5: Solution at time=1.0 Energy spectrum vs wavenumber for different final time of calculation for different values of $\nu_\alpha$ for fixed $\zeta$=0.01. a) $\nu_\alpha$=0.001 b) $\nu_\alpha$=0.01 c) $\nu_\alpha$=0.1 d) $\nu_\alpha$=0.5

With fractional order 1/3, the decay rate is almost identical to Kolmogorov's. This is not a coincidence as the Fourier symbol of the fractional Laplacian matches the scaling of the energy transfer time in Kolmogorov theory.

Classical Burgers shows a steeper decay, due to shock-dominated solutions. It dissipates energy at small scales too efficiently, unlike turbulence. Adding a fractional anti-diffusion term slows down dissipation at high, preserving energy in small scales. This makes the fractional model more suitable for mimicking energy cascades seen in real turbulence.

**(e) Simulation of Energy Decay in Turbulent Flows**

The decay of kinetic energy in turbulent flows is simulated by numerically solving the incompressible Navier-Stokes equations using a pseudo-spectral method in a three-dimensional periodic domain. The velocity field is initialized in Fourier space with divergence-free random components to ensure incompressibility. The spatial discretization employs a uniform grid of $64^3$ points, which provides a balance between computational feasibility and resolution of relevant flow scales; however, this moderate resolution may limit the capture of the smallest turbulent structures and the full inertial range dynamics. Time integration is performed using an explicit scheme with a fixed time step, advancing the velocity Fourier coefficients while enforcing the incompressibility constraint via a Helmholtz projection. Nonlinear advection terms are computed in physical space using inverse FFTs and then transformed back to Fourier space, leveraging the spectral accuracy of the method. Periodic boundary conditions are applied in all three spatial directions, simplifying the spectral formulation and mimicking an unbounded, homogeneous turbulence environment. Despite its high accuracy for smooth fields, the pseudo-spectral method can be sensitive to aliasing errors and requires dealiasing strategies, which were not explicitly addressed here, potentially affecting the fidelity of nonlinear term computations. The simulation also includes a fractional dissipation term characterized by a fractional Laplacian with order $\alpha=1/3$ to model anomalous diffusion effects, allowing comparison against classical viscous dissipation. Energy decay is monitored over time, and spectral energy distributions are analyzed to assess compliance with Kolmogorov scaling. Visualization of vorticity slices offers qualitative insight into flow structure evolution under classical and fractional dissipation regimes. Overall, this framework captures key features of turbulent energy decay while acknowledging the limitations inherent in numerical resolution and boundary idealizations.

Figure 7 demonstrate the effect of spatial fractional term for different final solution time.

The initial spectra of the classical NSE and the fNSE differ slightly because the fractional term immediately damps all modes, even at the first-time step.

The fractional Laplacian $(-\Delta)^{1/3}$ introduces non-local dissipation, acting across all scales (large and small), unlike classical viscosity $\nu\nabla^2$, which primarily damps energy at small scales (high wavenumbers k). This key difference explains two observed behaviors: The steep slope in E(t) (rapid decay) for the fractional case occurs because dissipation affects all scales simultaneously, including energy-containing large eddies.

In contrast, classical viscosity dissipates energy only at small scales, leading to slower decay as large-scale energy persists longer. The fractional term preserves the energy cascade, allowing a closer approximation of Kolmogorov −5/3 law in the inertial range. By dissipating energy more uniformly, it avoids the over damping of small scales that truncates the inertial range in classical NSE. Classical viscosity produces a sharp drop at high k (dissipation range dominance), while the fractional Laplacian yields a gentler dissipation, extending the −5/3 power law.

The observed differences in the energy spectra between the classical and fractional Navier-Stokes cases arise from their distinct dissipation mechanisms. In the classical case, the viscous term $\nu\nabla^2 u$ introduces dissipation scaling as $\sim \nu k^2$, which dominates sharply at high wavenumbers, truncating the inertial range and often leading to a bottleneck effect near the dissipation scale. In contrast, the fractional Laplacian $\nu_\alpha(-\Delta)^\alpha u$ (with $\alpha=1/3$) dissipates energy more gradually, scaling as $\sim \nu_\alpha k^{2/3}$. This weaker but broader dissipation preserves the Kolmogorov $k^{-5/3}$ inertial range over a wider span of scales, delaying the onset of the dissipation range and reducing spectral pile-up. Consequently, the fractional case exhibits a more extended segment aligning with the theoretical −5/3 slope, as its dissipation does not disrupt the energy cascade as abruptly as classical viscosity. This behavior highlights the fractional model's ability to better capture turbulence's multiscale dynamics, particularly in reproducing the inertial range scaling predicted by Kolmogorov theory. The classical case, while stable at our tested parameters, tends to over-damp high-k modes, shortening the inertial range, an effect that would become more pronounced at higher Reynolds numbers or resolutions. These spectral differences underscore the fractional operator's role in modeling turbulent flows with more realistic energy distribution across scales.

Figure 8 shows the influence of adding a Caputo derivative. For this benchmark problem the influence of temporal memory effect is negligible.

We conjecture that simulating on a refined grid would further improve the fractional case's agreement with Kolmogorov scaling, as higher resolution would better resolve the inertial range dynamics and mitigate numerical artifacts. However, this study is based on preliminary simulations with limited grid resolution, and the current results are insufficient to draw definitive conclusions about the physical behavior of the system. Due to constraints in grid spacing and potential numerical errors, we refrain from deeper analysis of the solution's physical aspects at this stage. These observations nevertheless highlight the fractional model's potential to better capture multiscale turbulent phenomena, pending more rigorous numerical validation in future work.

**Numerical Accuracy of Simulation**

The Viscous stability condition is

$\Delta t \lesssim 1/\nu k_{max}^2$, where $k_{max}=N/2$.

For $\nu=1e^{-5}$, N=64, $k_{max}=32$, this requires $\Delta t < 1/e^{-5} \times 1024 \approx 0.01$.

For the advective instability, the nonlinear term can feed energy into high-k modes faster than viscosity can dissipate it. The fractional case avoids this because the $\nu_\alpha=0.015$ term dominates over $\nu=1/e^{-5}$, providing stronger damping. The fractional operator smooths more aggressively across scales, preventing energy buildup at high k. The fractional term $(-\Delta)^\alpha$ has a regularizing effect. It acts like a low-pass filter with a smoother cutoff than classical viscosity. The effective damping rate is $\sim \nu_\alpha k^{(2\alpha)}$, which, for $\nu_\alpha=0.015$ and $\alpha=1/3$, is much larger than $\nu k^2$ at most scales.

a)

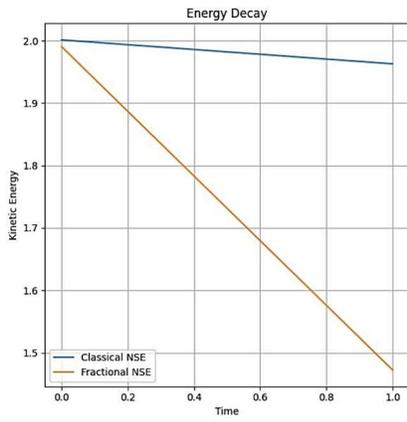
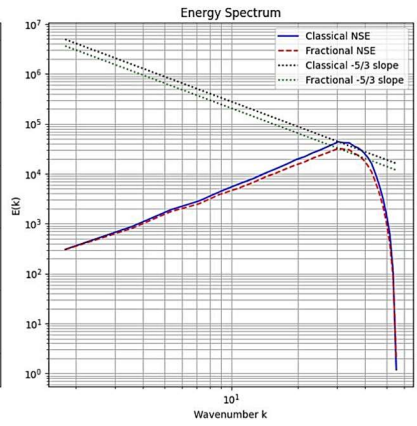
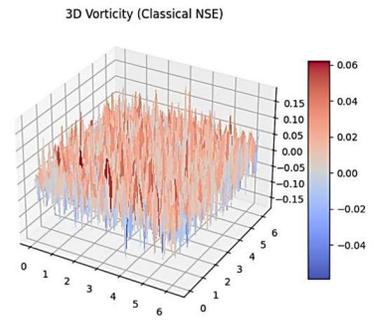
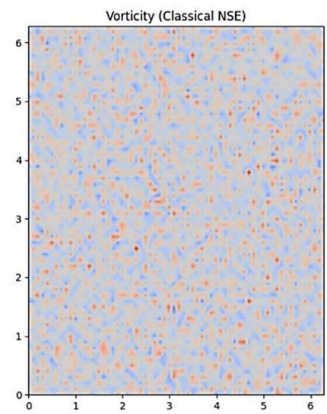
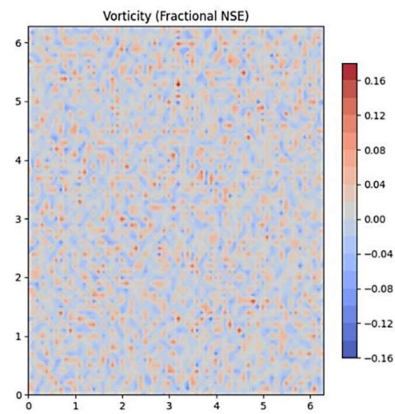
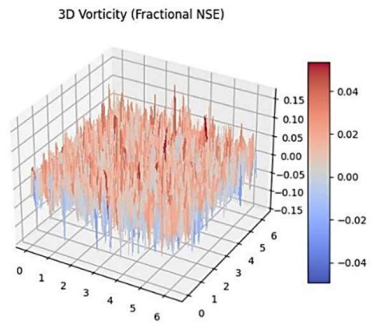

b)

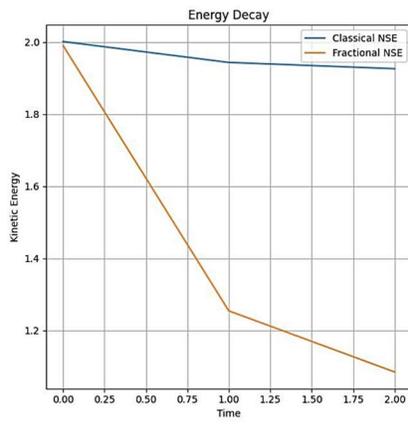
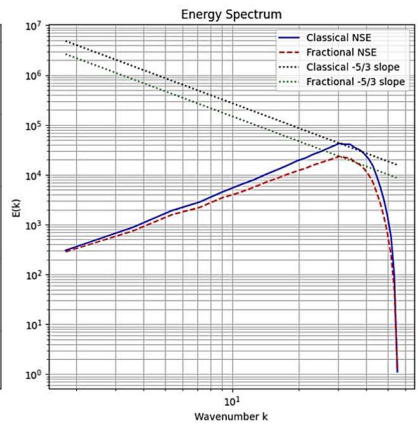
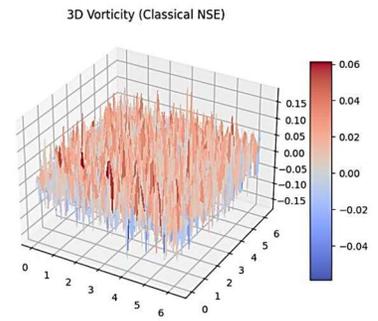
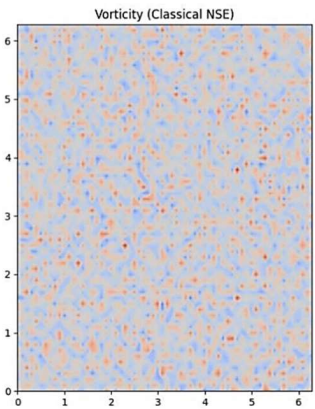
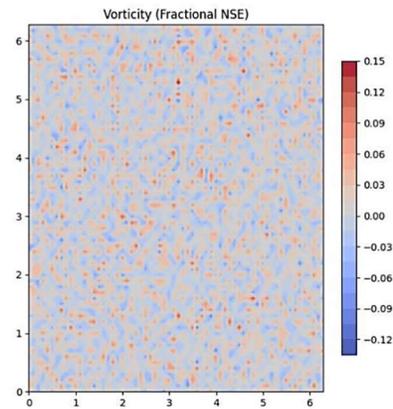
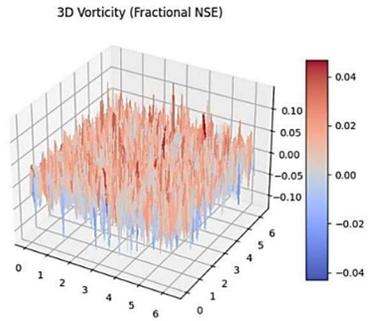

c)

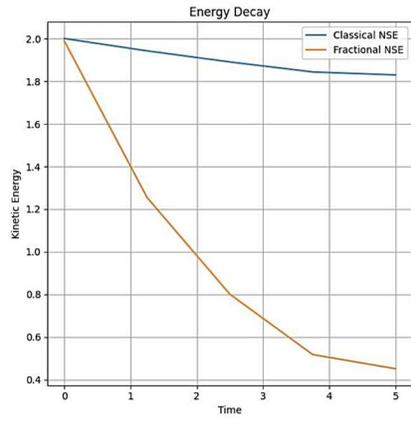 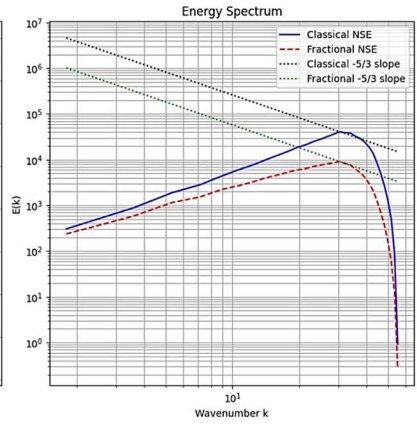 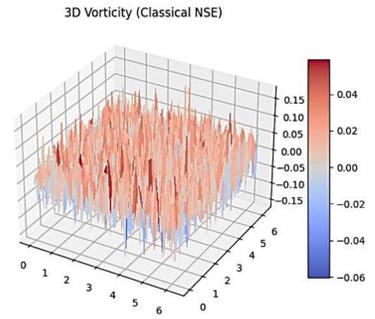

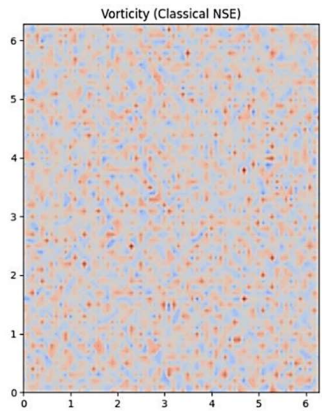 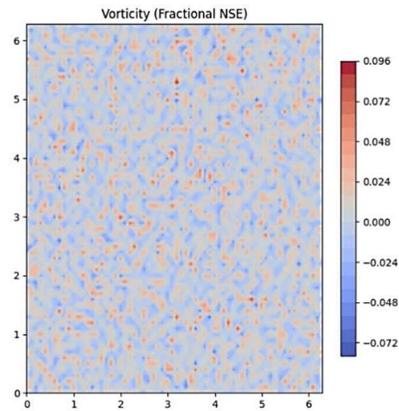 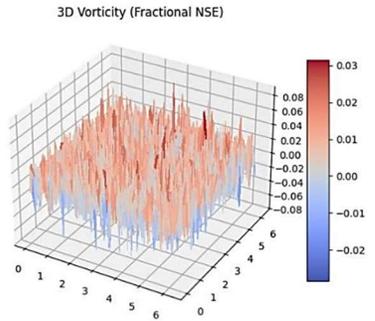

d)

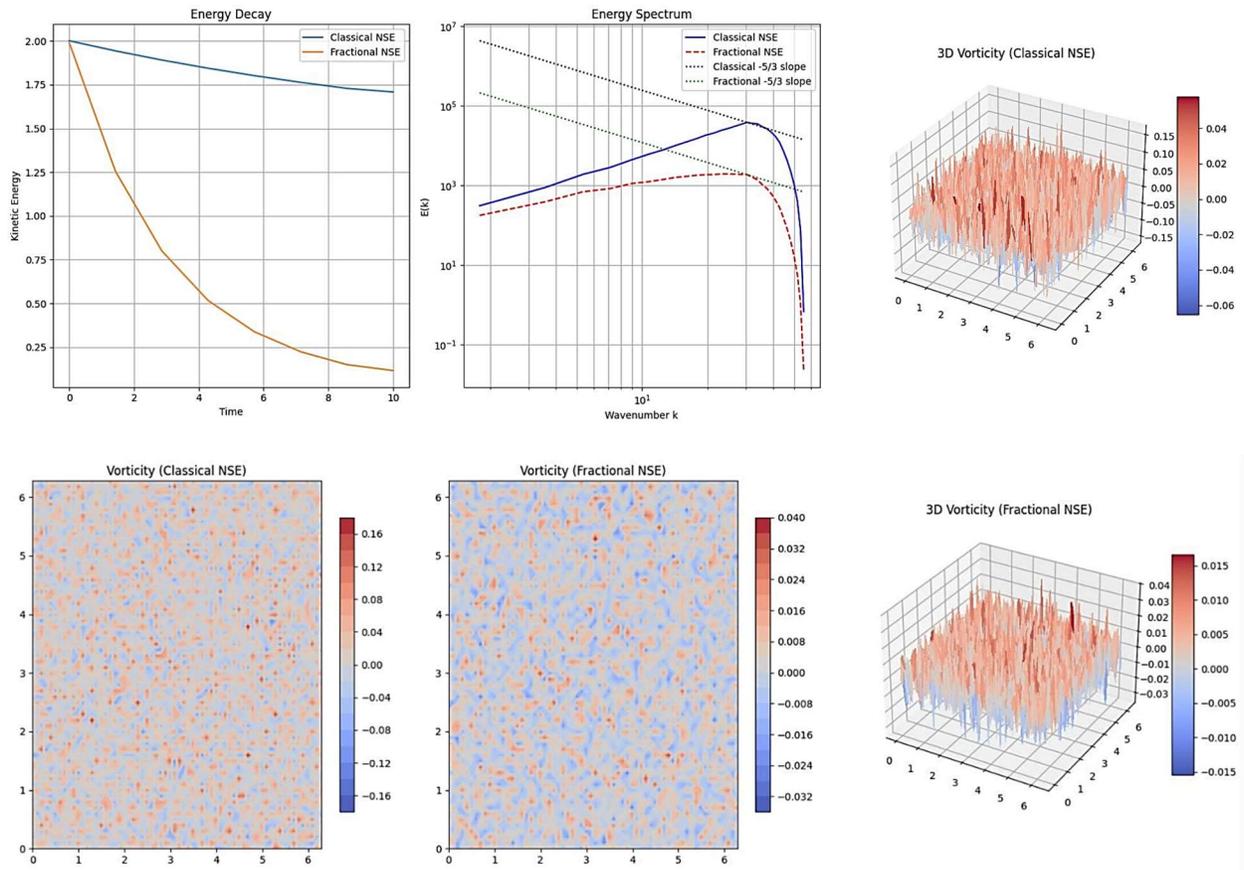

e)

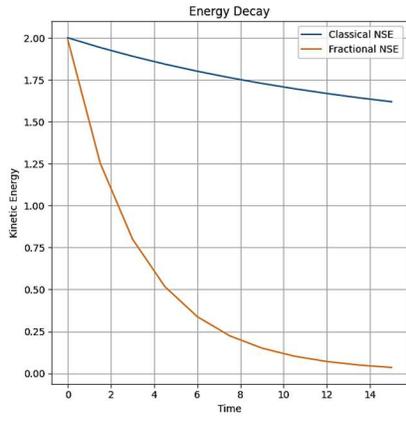
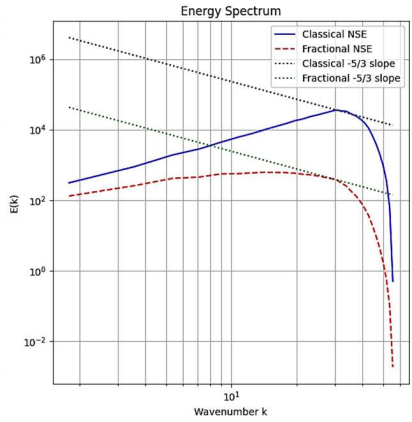
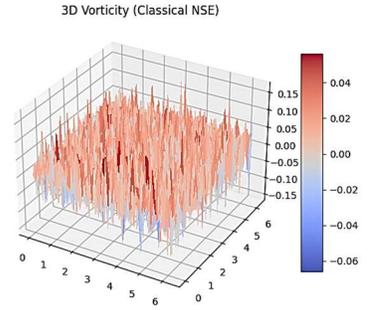
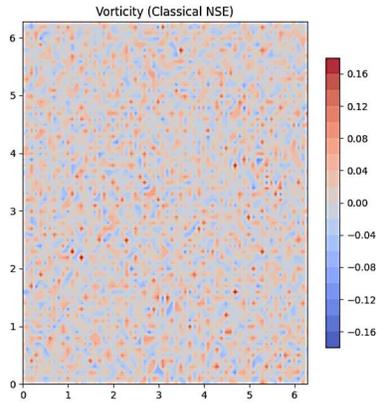
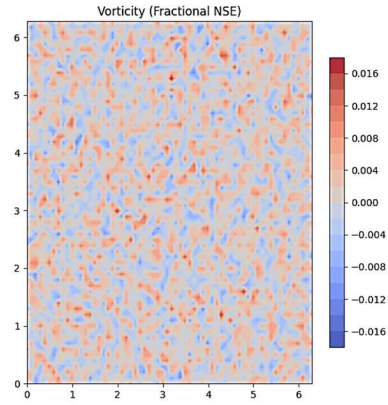
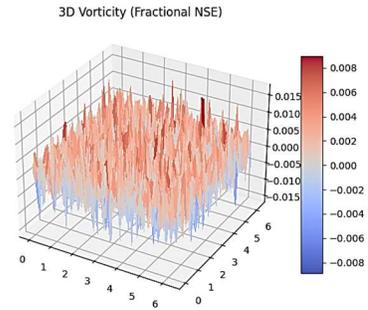

f)

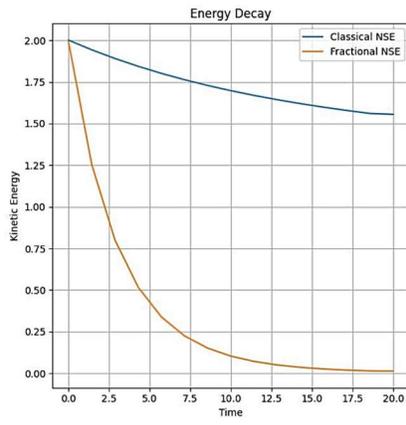 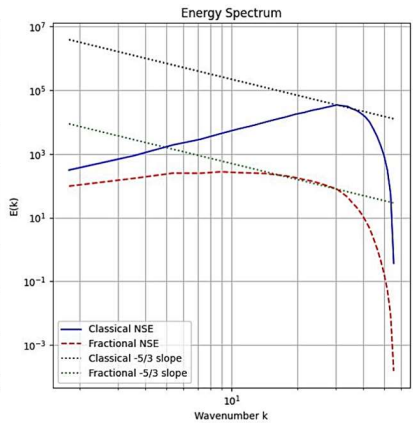 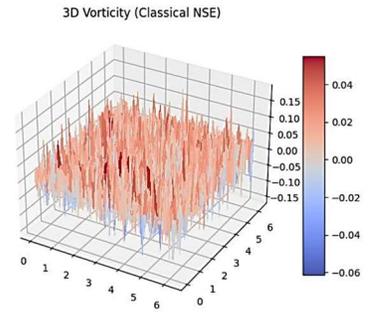

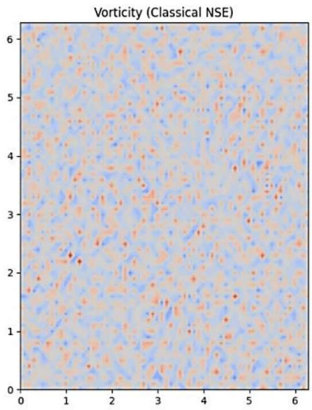 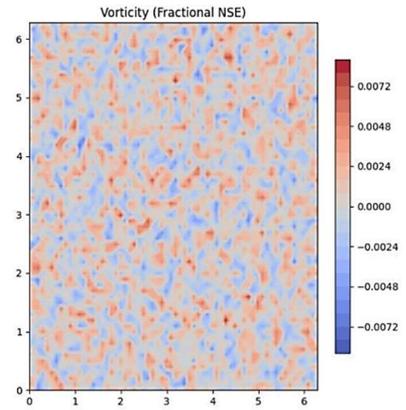 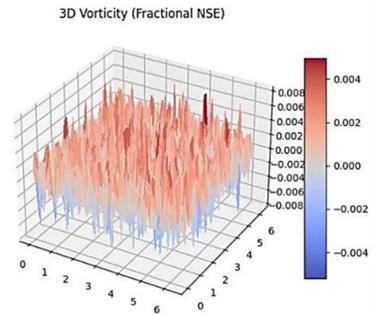

g)

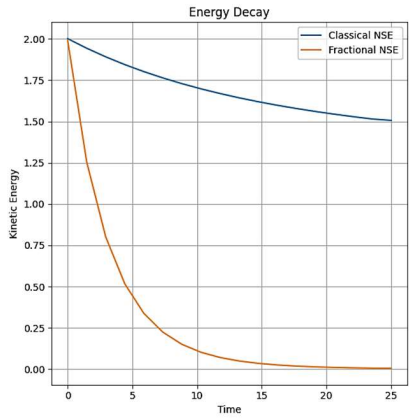
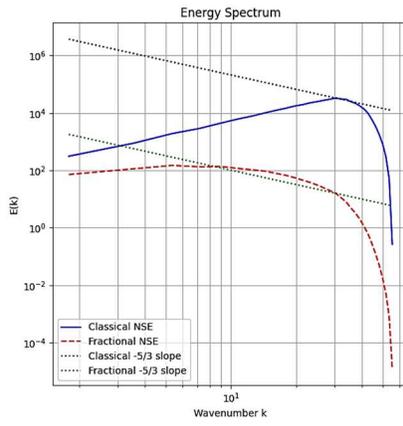
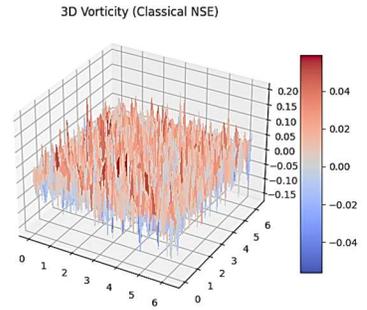
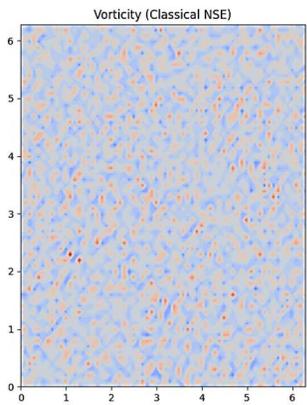
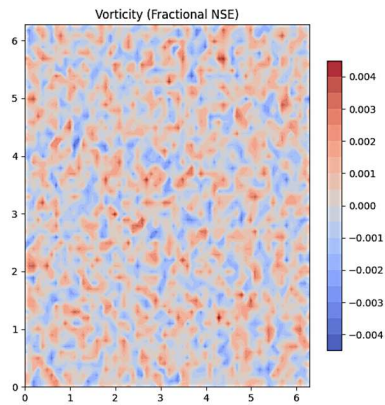
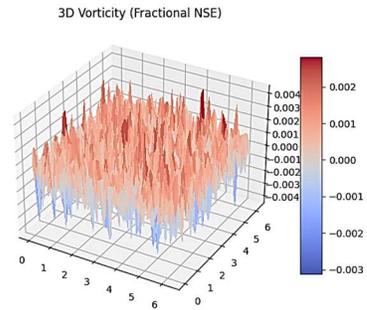

h)

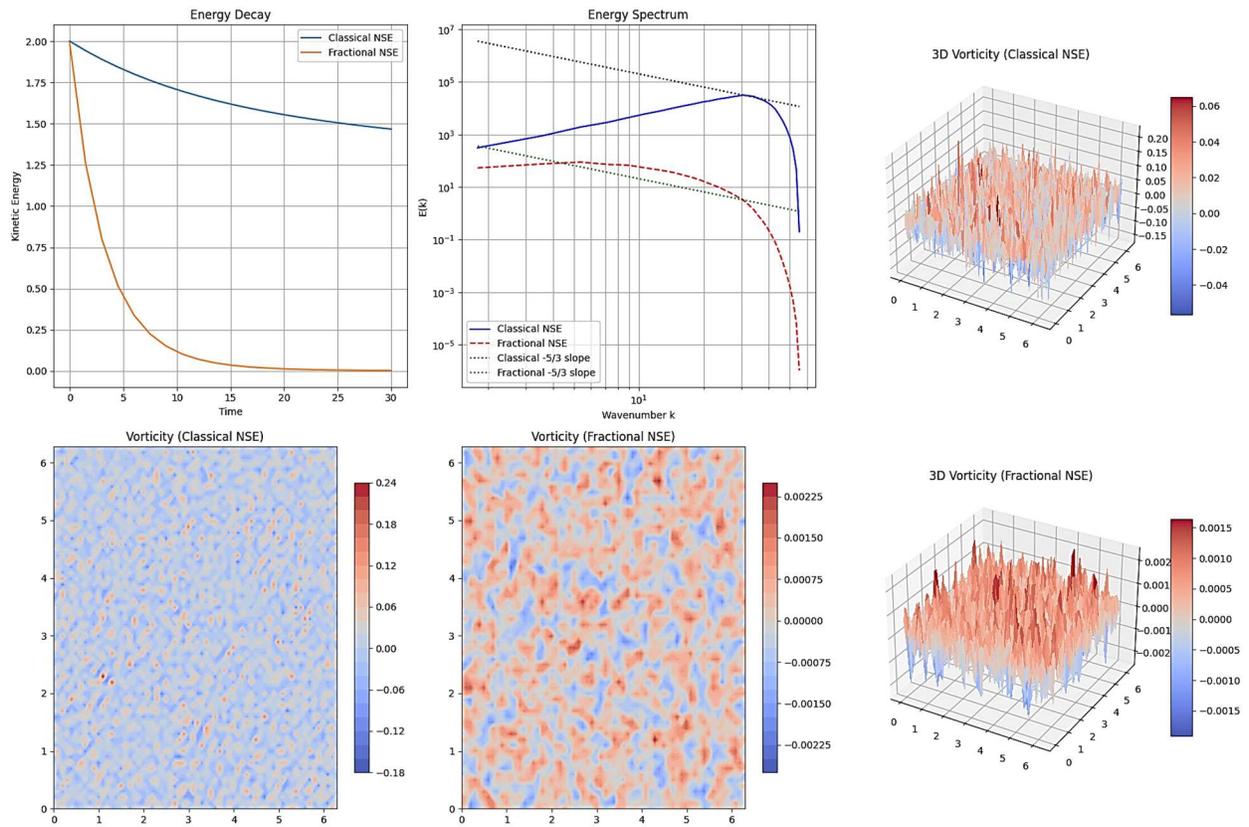

Fig 7: Turbulent kinetic energy vs time, Energy spectrum vs wavenumber, and vorticity contours for different final solution time. a) $T_{final}$ = 1.0 b) $T_{final}$ = 2.0 c) $T_{final}$ = 5.0 d) $T_{final}$ = 10.0 e) $T_{final}$ = 15.0 f) $T_{final}$ = 20.0 g) $T_{final}$ = 25.0 h) $T_{final}$ = 30.0

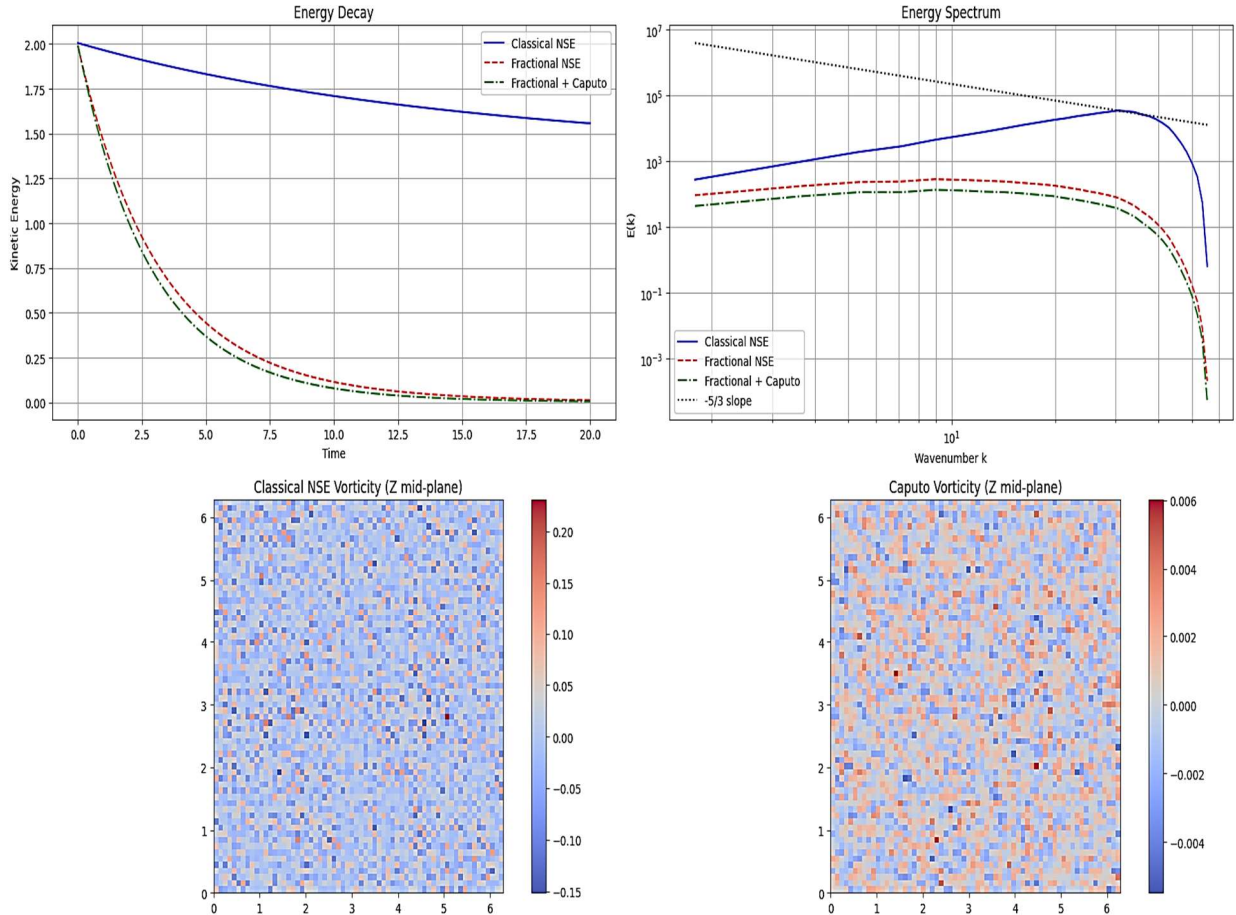

Fig 8: Turbulent kinetic energy vs time, Energy spectrum vs wavenumber, and vorticity contours for $\zeta=0.01$ at $T_{final}=20$

## 5. Conclusion

This study studies another model for turbulence flow simulation through a fractional generalization of the Navier–Stokes equations. By modifying the constitutive relation between stress and strain, rather than the governing equations directly, the fNSE presented in this paper preserves the structural integrity of classical fluid mechanics while incorporating key features of turbulent flow namely, nonlocal spatial interactions and memory effects.

Our derivation demonstrates that the fractional Laplacian of order $\alpha=1/3$ aligns with Kolmogorov's scaling and offers an inertial-range regularization without adjustable parameters. The fractional time derivative of order $\beta=1/2$, models physically plausible memory effects and may prove useful in near-dissipation regimes.

Ultimately, the fractional Navier–Stokes model may lead to improvement of classical local closure assumptions in turbulence. While computational and analytical challenges remain, we contend that conceptual progress must precede technical optimization. This work is intended as a preliminary step toward a physically interpretable simulation of turbulence based on fractional dynamics. Researchers worldwide are invited to apply the fNSE formulation presented in this paper in numerical studies to validate, refine, and expand upon the results, particularly regarding computational cost, implementation within different numerical frameworks, and comparisons with Direct Numerical Simulation (DNS) and experimental data.

Shifting to a philosophical perspective, we argue that discretizing a continuum equation inherently breaks the one-to-one correspondence with the original continuous system and the resulting discrete formulation embodies its own numerical properties and artifacts, distinct from those of the continuum model. As Bergson (1907) contended, time is fundamentally continuous; therefore, in "Bergsonian" standpoint, discretizing the time derivative in numerical schemes departs from the true mathematical and physical nature of temporal evolution. The Navier–Stokes equations are not analytically integrable; consequently, any numerical approach, including Direct Numerical Simulation (DNS) or other sophisticated methods can at best approximate their behavior, capturing only a projection or limited representation of the underlying dynamic. In "Heideggerian" terms no matter how advanced the science, there remains an irreducible residue, a haunting mystery. Heidegger would urge us to stay attuned to that resistance, not to overcome it, but to listen to it. That's where Being whispers and we, in turn, dance to its rhythm.

## Appendix: Overview of Fractional Derivatives in Continuum Modeling

Fractional derivatives generalize the concept of integer-order differentiation to non-integer orders, offering a natural framework for describing memory effects, hereditary phenomena, and scale-dependent dynamics in complex systems. Unlike classical derivatives, which are local operators, fractional derivatives are inherently nonlocal, incorporating contributions from a finite or infinite history in time or space.

**Temporal Fractional Derivatives**

In modeling memory and time-dependent relaxation, the most common formulations are the Caputo and Riemann–Liouville derivatives.

Caputo Derivative (of order $\beta \in (0,1)$) is as follows:

$$Dt^\beta f(t) = \frac{1}{\Gamma(1-\beta)} \int_0^t \frac{f'(s)}{(t-s)^\beta} ds \qquad (A.1)$$

This form is often preferred in physical models because it allows for initial conditions in the classical form, i.e., f(0), f'(0), etc., which have physical interpretations.

The kernel (t−s)$^{-\beta}$ implies that recent history contributes more than distant history, but all past times s<t affect the present. This captures non-Markovian effects in time evolution.

**Spatial Fractional Derivatives and Operators**

Fractional spatial derivatives generalize classical diffusion or Laplacian operators to model nonlocal spatial interactions or anomalous diffusion.

Fractional Laplacian $(-\Delta)^{\alpha/2}$ defined via the Fourier transform:

$$F\left[(-\Delta)^{\frac{\alpha}{2}} u(x)\right](\xi) = |\xi|^\alpha \hat{u}(\xi) \qquad (A.2)$$

for $\alpha \in (0,2)$. This definition makes the operator pseudo-differential and nonlocal.

Equivalent Integral Form (in $R^d$) is expressed as:

$$(-\Delta)^{\frac{\alpha}{2}} u(x) = C_{d,\alpha}\, P.V. \int_{R^d} \frac{u(x) - u(y)}{|x-y|^{d+\alpha}}\, dy \qquad (A.3)$$

where P.V. denotes the Cauchy principal value. This form highlights the hypersingular nature of the operator and its connection to Lévy flights and long-range interactions.

# Acknowledgment

The authors gratefully acknowledge the assistance of AI-based language tools in improving the clarity and fluency of the English writing. All ideas, arguments, and scientific content are solely the responsibility of the authors.

## Statements and Declarations

No funding was received to assist with the preparation of this manuscript.